\documentclass[twocolumn,twocolappendix]{aastex631}

	\usepackage{graphicx}	
	\usepackage{amsmath}
	\usepackage{amssymb}	
	\usepackage{longtable}
	\usepackage{threeparttable}
	\usepackage{multirow}
	\usepackage{booktabs}
	\usepackage{array}
    \usepackage[figuresright]{rotating}
    \usepackage[T1]{fontenc}

	\shorttitle{sGRB with or without extended emission}
	\shortauthors{Li et al.}

	\graphicspath{{./}{figures/}}

\begin{document}
			\title{Comparison of the origin of Short Gamma ray Bursts with or without extended emission}


\correspondingauthor{Qi-Bin Sun}
\email{sunqibin@ynu.edu.cn}	
\author{Qin-Mei Li}
\affiliation{Department of Astronomy, School of Physics and Astronomy, Yunnan University, Kunming 650091, PR of China}
\affiliation{Guizhou University, Department of Physics, college of physics, Guizhou university, Guiyang, 550025, PR of China}

\author{Qi-Bin Sun}
\affiliation{Department of Astronomy, School of Physics and Astronomy, Yunnan University, Kunming 650091, PR of China}


\begin{abstract}
The merger of compact binary stars produces short gamma-ray bursts (sGRBs), involving channels such as neutron star - neutron star (BNS) and neutron star - black hole (NS-BH). The association between sGRB 170817A and gravitational wave GW 170817 provides reliable evidence for the BNS channel. Some speculations suggest that sGRBs with extended emission (EE) may represent another distinct population. The offset is the distance between the GRB sky localization and the host galaxy center. We compared the offset distributions of these two types of samples (46 sGRBs with EE and 9 without EE samples) and found that they follow the same distribution. Utilizing non-parametric methods, we examined the luminosity function and formation rate of sGRBs without any assuming.
The luminosity function can be described as $\psi(L_{0}) \propto L_{0}^{-0.12 \pm 0.01}$ for $L_{0} < L_0^b$ ($\psi(L_{0}) \propto L_{0}^{-0.73 \pm 0.02}$ for $L_{0} > L_0^b$) for sGRB without EE and $\psi(L_{0}) \propto L_{0}^{-0.13 \pm 0.003}$ for $L_{0} < L_0^b$ ($\psi(L_{0}) \propto L_{0}^{-0.61 \pm 0.01}$ for $L_{0} > L_0^b$) for sGRBs with EE. The formation rate is characterized as $\rho(z) \propto (1 + z)^{-3.04 \pm 0.10}$ for $z < 1$ and $\rho(z) \propto (1 + z)^{-0.29 \pm 0.38}$ for $1 < z < 3$ for sGRB without EE, while for sGRBs with EE, it is $\rho(z) \propto (1 + z)^{-3.85 \pm 0.15}$ for $z < 1$ and $\rho(z) \propto (1 + z)^{-0.40 \pm 1.11}$ for $1 < z < 3$.
Our findings suggest no significant difference in the progenitors of sGRBs with and without EE when considered in terms of spatial offsets, formation rates, and luminosity function.

\end{abstract}



	\section{Introduction} \label{sec:intro}
Gamma-ray bursts (GRBs) are brief, intensive flashes in the universe, produced in ultra relativistic jets \citep{2007ChJAA...7....1Z}. In the internal shock model, the central engine produces shells with comparable energy but different Lorentz factors. The slower shell, followed by a faster one, catches up and collides, generating observed GRBs \citep{1993ApJ...405..278M,1999astro.ph..9299F,1999MNRAS.309L..13L}. Traditionally, GRBs are divided into two categories based on $T_{90}$: short GRBs (sGRBs, $T_{90}<2$ s) typically exhibit higher spectral hardness compared to long GRBs (lGRBs, $T_{90}>2$ s) \citep{1993ApJ...413L.101K}.

The association between lGRBs and supernovae confirms that lGRBs originate from the deaths of massive stars \citep{1999A&AS..138..465G,2003ApJ...591L..17S,2024A&A...684A.164K}. sGRBs are produced by the mergers of binary compact objects: neutron star-neutron star (NS-NS) or neutron star-black hole (NS-BH) \citep{1989Natur.340..126E,2002ApJ...570..252P,2011ApJ...732L...6R}. Direct evidence for a binary objects merger includes the detection of an infrared excess observed in sGRB 130603B, interpreted as a kilonova resulting from the radioactive decay of unstable heavy elements like $^{56}Ni$ during rapid neutron capture process ($r$-process) \citep{2013Natur.500..547T,2013ApJ...774L..23B}. Increasingly, GRBs associated with kilonovae are being identified \citep{2015ApJ...811L..22J,2023MNRAS.521..269Z,2023ApJ...943..104Z,2023MNRAS.524.1096L}.

In 2017, the Advanced Laser Interferometer Gravitational-wave Observatory (aLIGO) and Advanced Virgo gravitational wave (GW) observatories detected a GW signal (GW 170817, \citealp{2017PhRvL.119p1101A}), and subsequently, the Gamma-ray Burst Monitor (GBM; \citealp{2009ApJ...702..791M}) aboard the Fermi satellite detected sGRB 170817A \citep{2017ApJ...848L..12A}, whose chirp was consistent with the merger of binary neutron stars (BNS). For NS-BH mergers, the neutron star is gradually disrupted by the black hole, and part of the neutron star's mass may remain outside the black hole, forming an accretion disk \citep{2005MNRAS.356...54D,2006PhRvD..74l1503S,2008PhRvD..77h4015S,2008ApJ...680.1326R}. In theory, both compact binary mergers (BNS or NS-BH) can produce sGRBs \citep{2005MNRAS.356...54D}.

Simulations have suggested that NS–BH mergers may disrupt the neutron star over several passages, resulting in the ejection of some material to large radii \citep{2005ApJ...634.1202R,2013ApJ...778L..16H}. Additionally, \citet{2020ApJ...895...58G} anticipated that the timescale for NS–BH mergers is longer. They also noted that the potential range of mass ratios in NS–BH mergers may differ from BNS mergers, which contributes to this longer timescale.
\citet{2008MNRAS.385L..10T} suggested that sGRBs with extended emission (EE) may originate from NS-BH mergers, based on offset comparisons due to kicks during mergers resulting in small average offsets from their host galaxies \citep{2003Sci...300.1119M,2006ApJ...648.1110B}, and sGRBs are less likely to have optically detected afterglows, consistent with BNS mergers occurring in low-density environments. However, \citet{2013ApJ...776...18F} found offsets of sGRBs with and without EE drawn from the same distribution, indicating no evidence to confirm they are produced in different progenitor systems (4 sGRB EE). \citet{2020ApJ...895...58G} provided evidence supporting the hypothesis that sGRB EE constitutes the NS-BH merger population by comparing offsets and energies. However, it is currently unclear whether NS-BH should be connected to a specific offset distribution. \citet{2022ApJ...936L..10Z} found that the EE of GRB 211211A could result from the fallback accretion of r-process heating materials, predicted to occur after NS-BH mergers. Model fitting results indicate that the NS-BH merger rate is higher than that of BNS mergers, considering scenarios where the neutron star is disrupted and not entirely swallowed by the black hole \citep{2014PhRvD..90b4026F,2017CQGra..34d4002F}. \citet{2018MNRAS.479.4391M} found different merger rates for BNS and NS-BH mergers, depending on assumptions about the efficiency of common envelope (CE) ejection. Using the formation rate (FR) of the sGRBs to speculate on the merger rate of BNS and NS-BH is a common approach (eg.,\citealp{2004JCAP...06..007A,2018ApJ...852....1Z}). Based on the assumption that sGRB without EE and sGRB EE are connected to the BNS and NS-BH merger, respectively. We compared their offset, FR and luminosity function (LF) to identify whether their origins are the same.

In this paper, our main purpose is to study the LF and FR of sGRBs with or without EE using Lynden-Bell's c$^{-}$ method to distinguish their origin. Section \ref{sec:DATA} provides a detailed introduction to the sample sources, the K-correction method, the Yonetoku relation, Lynden-Bell's c$^{-}$ method, and the Kendall $\tau$ test methods. 
In Section \ref{sec:luminosityfunction}, we present the results of the LF and FR of sGRBs (with and without EE). Finally, Section \ref{sec:DISCUSSION} presents the conclusions and discussions.


\section{DATA and k-correction} \label{sec:DATA}

\begin{figure*}
\centering
\includegraphics[width=\columnwidth]{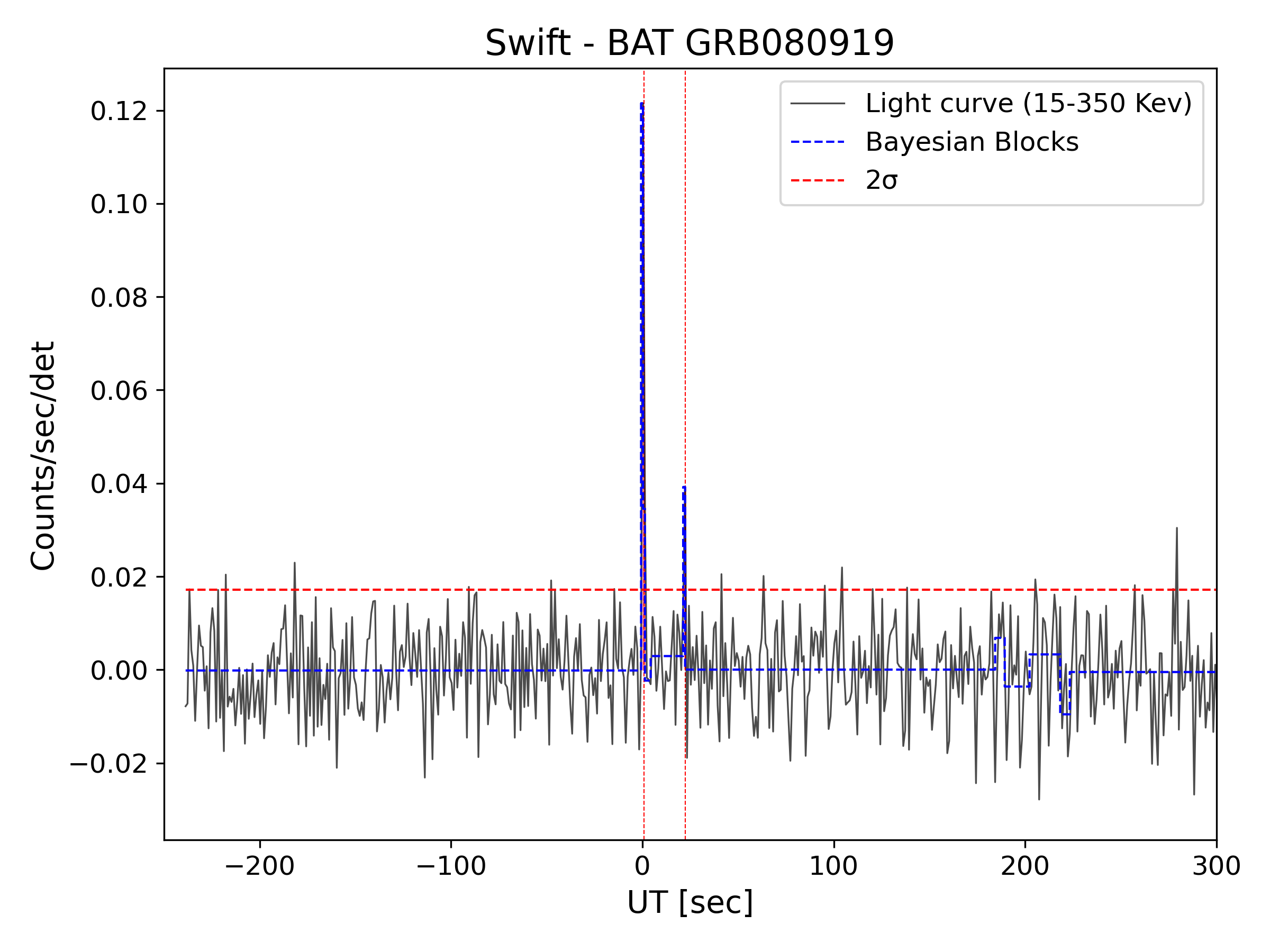}
	\includegraphics[width=\columnwidth]{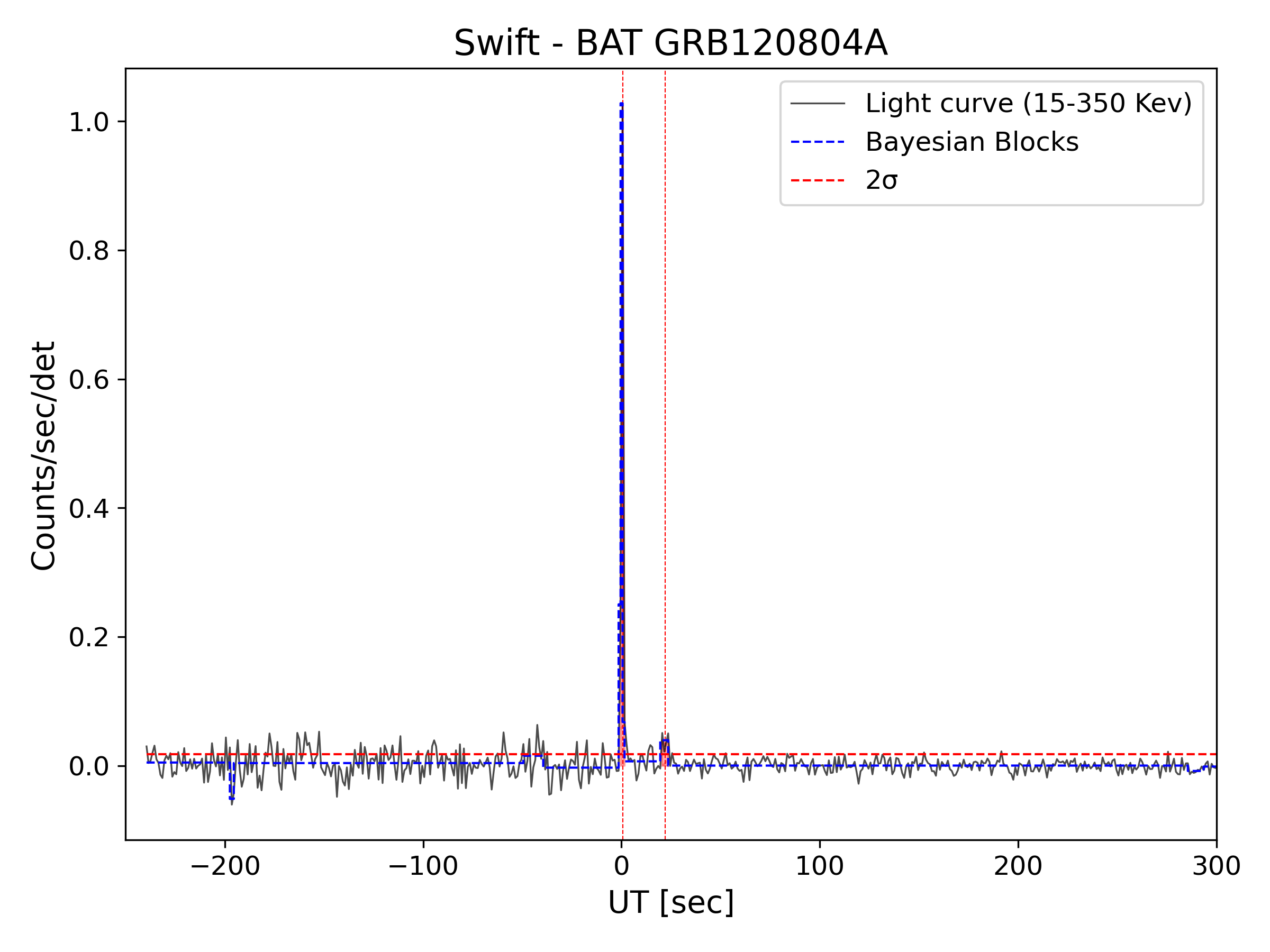}
	\caption{Typical example of sGRB with extended emission. The solid blue line is the statistical result of the Bayesian block method for the light curve. The vertical red dashed line represents that this component is greater than 2 $\sigma$. The horizontal red dashed line refers to 2 $\sigma$.}
	\label{fig:9}
\end{figure*}

In this paper, we use data provided by the Neil Gehrels Swift Observatory \citep{2004NewAR..48..431G} \footnote{\url{https://swift.gsfc.nasa.gov/archive/grb_table/}}. Swift is a first-of-its-kind multi-wavelength observatory dedicated to the study of GRB science. Since 2004, Swift has discovered approximately 1650 GRBs, including 138 sGRBs. Two sGRBs lacked a light curve, the spectra of some sGRBs could not be successfully fitted using the cut-off power law model \citep{2008ApJS..175..179S}. Additionally, some samples had missing peak flux values, resulting in a final dataset of 95 samples. 
The Burst Alert Telescope (BAT) detects GRBs and accurately determines their positions in the sky. This instrument provides light curves in different energy channels in the $\gamma$-ray band, including Channel 1 (15–25 keV), Channel 2 (25–50 keV), Channel 3 (50–100 keV), Channel 4 (100–350 keV), and Channel 5 (15–350 keV). Figure \ref{fig:9} presents two typical examples of GRBs with EE components: GRB 080919A and GRB 120804A.

The Bayesian Block Algorithm, introduced by \citet{2005ISPL...12..105J} and refined by \citet{2013ApJ...764..167S}, is a powerful statistical tool for detecting change points in time series data. It segments the data into intervals or "blocks", each with a constant rate or flux, optimizing these segments using Bayesian principles. This approach handles statistical uncertainties and noise effectively, making it ideal for irregular events such as GRBs and pulsar light curves.
Widely used in GRB research \citep{2014MNRAS.445.2589B,2019ApJ...876...89V,2024MNRAS.527.7111L}, the algorithm excels at revealing significant changes and patterns in complex datasets, offering insights into underlying emission mechanisms. 
Since the extended emission (EE) component often occurs in low-energy channels, we use a signal-to-noise ratio (S/N $>$ 2$\sigma$) \citep{2015MNRAS.452..824K,2020RAA....20..201Z} and the Bayesian block method to identify EE components from Channels 1, 2, 1+2, and 5.

Ultimately, we selected 95 Swift sGRBs (65 sGRBs with EE and 30 sGRBs without EE). Unfortunately, most sGRBs do not have well-estimated redshifts. Therefore, this article adopts empirical relationships to estimate redshifts. The empirical relationships are established using the 22 \textit{Swift} sGRBs samples presented in Table 1 of \cite{2021RAA....21..254L} and an additional 4 \textit{Swift} sGRBs from \citet{2024MNRAS.527.7111L}.

The spectra of \textit{Swift} GRB are generally fitted by two spectral models, a single/cut-off power-law model (CPL; \citealp{2008ApJS..175..179S}). The form of the CPL function is as follows:
\begin{equation}
	\Phi (E) = B{\left( {\frac{E}{{50keV}}} \right)^\alpha }\exp \left( { - \frac{{(2 + \alpha )E}}{{{E_P}}}} \right)
\end{equation}
 Table \ref{tab:1} and \ref{tab:2} list information for sGRBs including the GRB name, duration $T_{90}$, redshift $z$, photon index $\alpha$, peak energy $E_p$ in the observer frame, peak flux $P$, and bolometric peak luminosity $L_p$.
Since peak flux is observed in different energy ranges, we will use the same K-correction method to convert the flux into the $1-10^5$ keV band to get bolometric luminosity (eg., \citealp{2015ApJS..218...13Y}). The bolometric luminosity can be calculated by $L = 4\pi D_L^2(z)PK$. $K$ and $P$ are the K-correction factor and the peak flux observed in the energy range, respectively. If the P is in units of $erg/c{m^2}/s$, the K can be expressed as
\begin{equation}
	K = \frac{{\int_{1/(1 + z)}^{10000/(1 + z)} {E\Phi (E)dE} }}{{\int_{{E_{\min }}}^{{E_{\max }}} {E\Phi (E)dE} }}
\end{equation}
else if the P is in unit of $photon/c{m^2}/s$,  the K can be expressed as
\begin{equation}
	K = \frac{{\int_{1/(1 + z)}^{10000/(1 + z)} {E\Phi (E)dE} }}{{\int_{{E_{\min }}}^{{E_{\max }}} {\Phi (E)dE} }}
\end{equation}
the flux limit is of chosen as the observation limited line of \textit{Swift} for sGRBs without EE (${F_{\lim}} = 1.0 \times {10^{ - 8}}erg/c{m^2}/s$; \citealp{2008MNRAS.388.1487L}). To more tightly close the sample and keep sample above the luminosity limit, the flux limit of sGRBs with EE is set as ${F_{\lim}} = 2.0 \times {10^{ - 8}}erg/c{m^2}/s$. Then, the luminosity limit is given as ${L_{\lim}} = 4\pi D_L^2(z){F_{\lim}}$ within 10-150 keV energy range.

\subsection{offset  \label{sec: method}}
\begin{figure}
	\centerline{\includegraphics[width=\columnwidth]{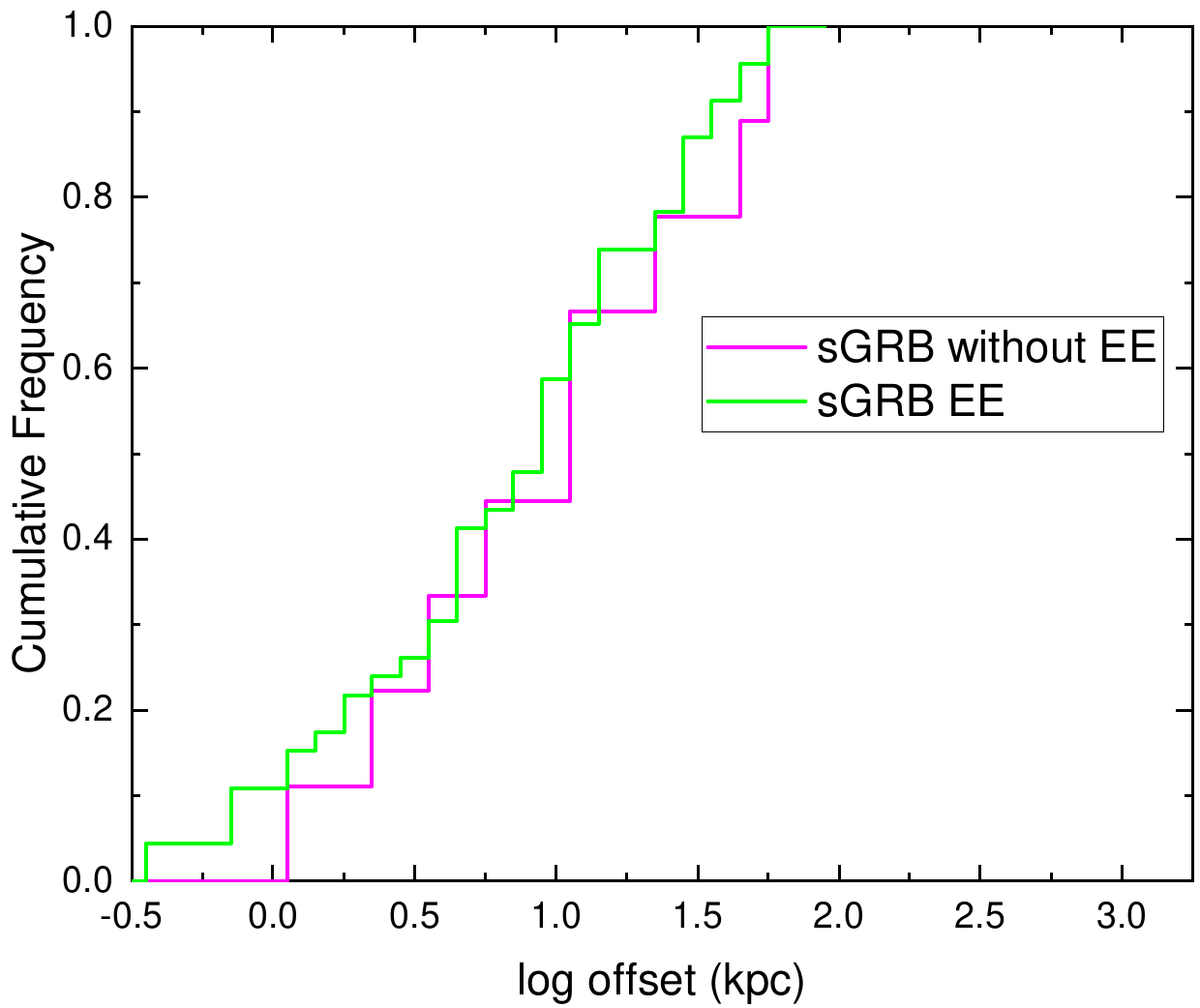}}
	\caption{Cumulative frequency distribution of offset for sGRB with (green line) and without EE (purple line).}
	\label{fig:10}
\end{figure}

Offset is the distance between the GRB sky localization and the host galaxy center. We obtain the offset of a portion of the sample from the literature \citep{2022ApJ...940...56F,2022MNRAS.515.4890O,2023ApJ...950...30Z}. Figure \ref{fig:10} represents the cumulative frequency distribution of logarithms offset for sGRBs with or without EE (9 sGRB without EE and 46 sGRB EE). The mean value of offset for sGRB without EE and sGRB EE are 12.88 kpc and 10.23 kpc. We apply a Kolmogorov-Smirnov (KS) test to the offset distribution of sGRBs with and without EE. We get the statistic $D=0.16$, which is less than the critical value of $D(n_1,n_2)=0.50$ for $n_1=9$ and $n_2=46$, signifying that their offset share the same distribution.

Our result is consistent with \citet{2013ApJ...776...18F} (16 sGRB without EE, 5.3 kpc; 4 sGRB EE, 3.2 kpc ), indicating no evidence to confirm they are produced in different progenitor systems. Our result have a large difference with \citet{2008MNRAS.385L..10T}, who found the mean value of 5 sGRB EE is 3.95 kpc and the mean offset of sGRB without EE is 31.9 kpc. The reason for this discrepancy is that their short hard GRB EE samples include events with $T_{90} > 2\,\text{s}$ detected by the BAT instrument (for example, 050724 with $T_{90} = 152\,\text{s}$; 061210 with $T_{90} = 85\,\text{s}$), while we selected samples with $T_{90} < 2\,\text{s}$, which falls outside our statistical range. \citet{2020ApJ...895...58G} found that the distributions of offset for non-collapsar and EE samples are statistically distinct, with EE bursts exhibiting high energies and low offsets. This trend aligns with the expected characteristics of NS-BH mergers, leading them to suggest that sGRB EE are a result of such mergers.

\subsection{Yonetoku relation  \label{sec: method}}
We fit the $E_{p}$-$L_p$ correlation \citep{2004ApJ...609..935Y} with the linear form
\begin{equation}
	\log (Lp) = a + b\log E_{p,i}^{}\
\end{equation}
The best fit of this correlation is given by 
\begin{equation}
	\log (Lp) = ({\rm{47}}{\rm{.30}} \pm {\rm{0}}{\rm{.59)}} + ({\rm{1}}{\rm{.58}} \pm {\rm{0}}{\rm{.21)}}\log E_{p,i}^{}
\end{equation}

\begin{figure}
	\centerline{\includegraphics[width=\columnwidth]{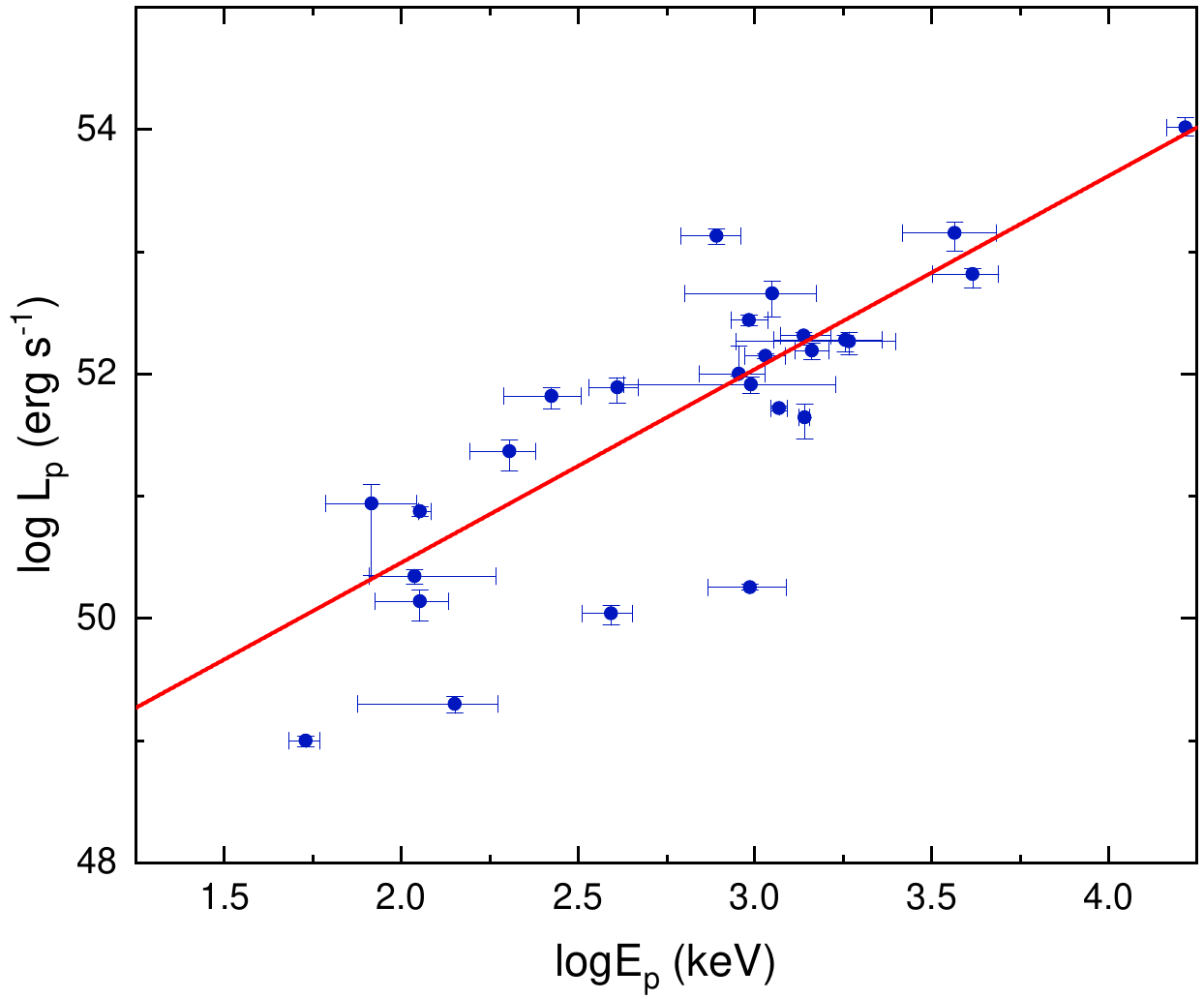}}
	\caption{The $E_{p}$-$L_p$ correlation for sGRBs. The red line is the result of best fit.}
	\label{fig:1}
\end{figure} 
where Pearson's correlation coefficient is 0.84, which is consistent with the correlation obtained by \citet{2018ApJ...852....1Z}. Figure \ref{fig:1} shows the $E_{p}$-$L_p$ correlation. The red line is the best-fitting result. We can rewrite the Equation (5) as
\begin{equation}
	\frac{{D_l^2}}{{{{(1 + z)}^{1.58}}}} = \frac{{1.99 \times {{10}^{47}}}}{{4\pi {F_p}}}E_{p,i}^{1.58}
\end{equation}
The pseudo redshifts can be derived using the observed peak energy $E_{p,obs}$ according to the above equation and peak flux $F_p$.

\subsection{Lynden-Bell's ${c^ - }$ method} \label{sec:c- method}
Lynden-Bell's ${c^ - }$ method is an efficient method for a truncated data sample. This method is used on a variety of cosmological objects, such as quasar \citep{1971MNRAS.155...95L,1992ApJ...399..345E}, GRB \citep{2015ApJS..218...13Y,2024MNRAS.527.7111L}, and FRB \citep{2019MNRAS.487.3672Z,2024arXiv240603672C}. In this paper, we also use this method to drive the cosmological evolution of the LF and FR of our sample.

\begin{figure*}
\centering
	\includegraphics[width=\columnwidth]{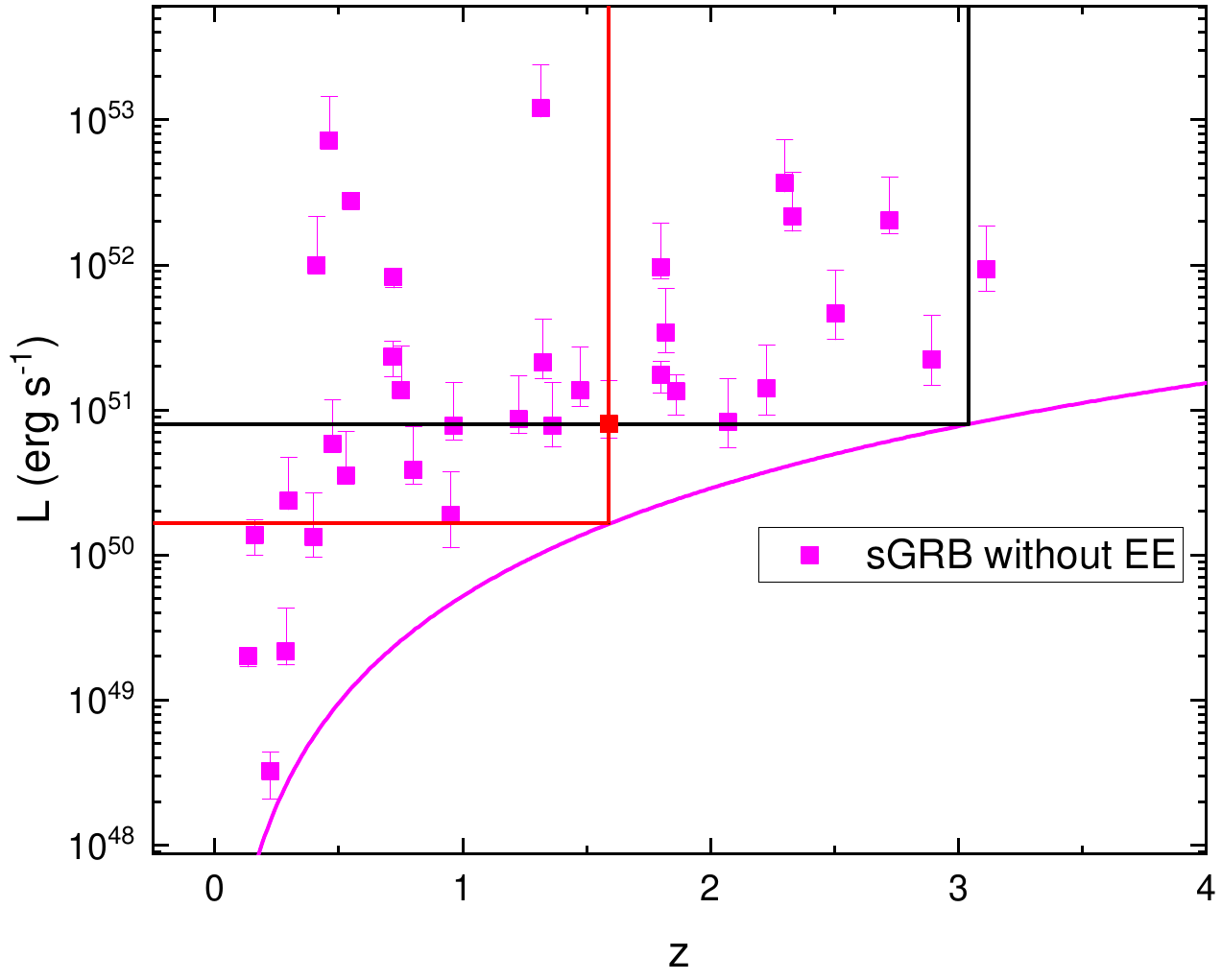}
	\includegraphics[width=\columnwidth]{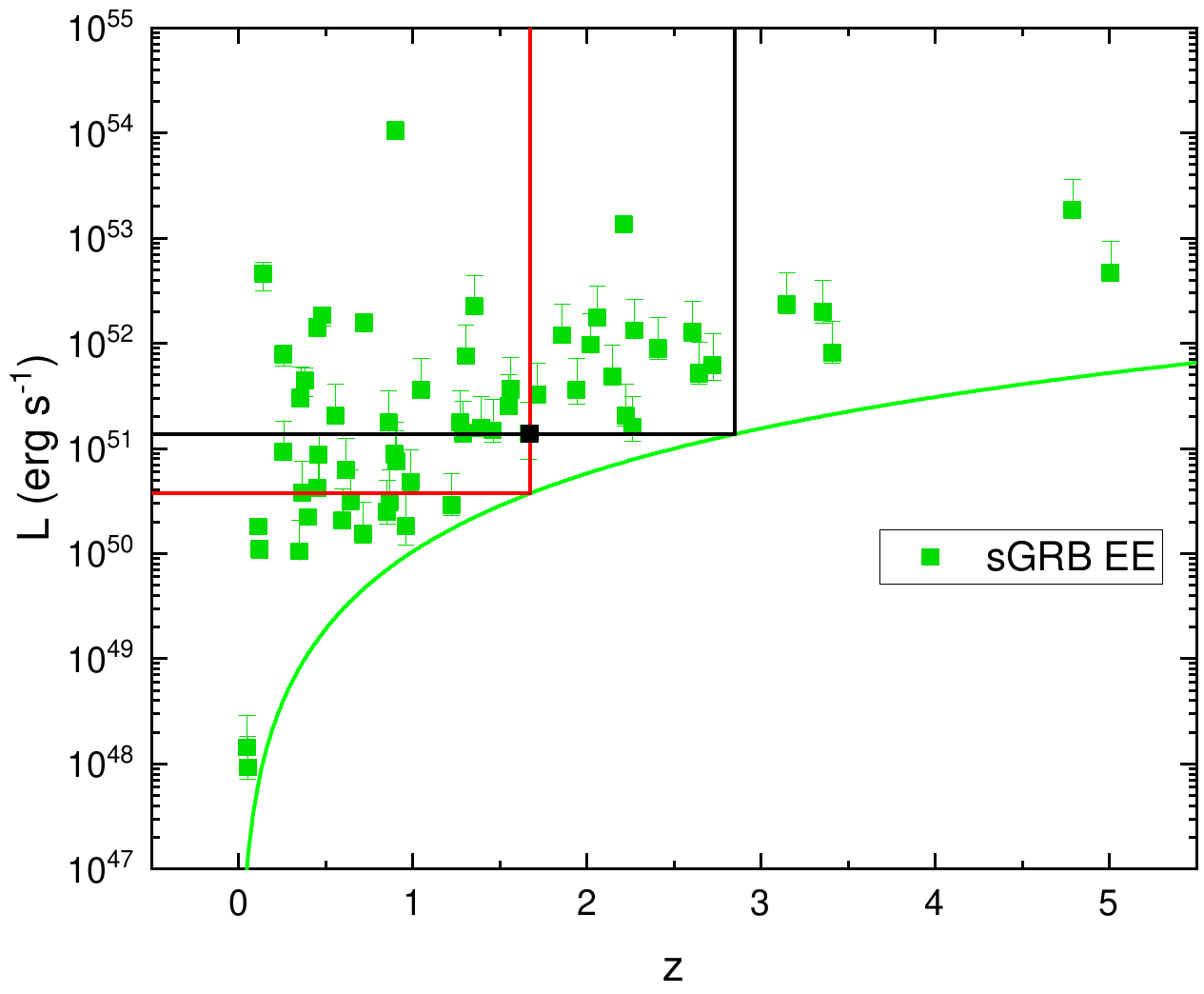}
	\caption{The redshift distributions of sGRB without EE (left panel) and with EE (right panel), estimated using the $E_{p}$-$L_p$ correlation, are depicted. In the left panel, the purple line represents a flux limit of $1 \times 10^{-8} \text{ erg}, \text{cm}^{-2} \text{s}^{-1}$, while in the right panel, the green line denotes a flux limit of $2 \times 10^{-8} \text{ erg}, \text{cm}^{-2} \text{s}^{-1}$.}
	\label{fig:4}
\end{figure*}

\begin{figure*}
\centering
	\includegraphics[width=\columnwidth]{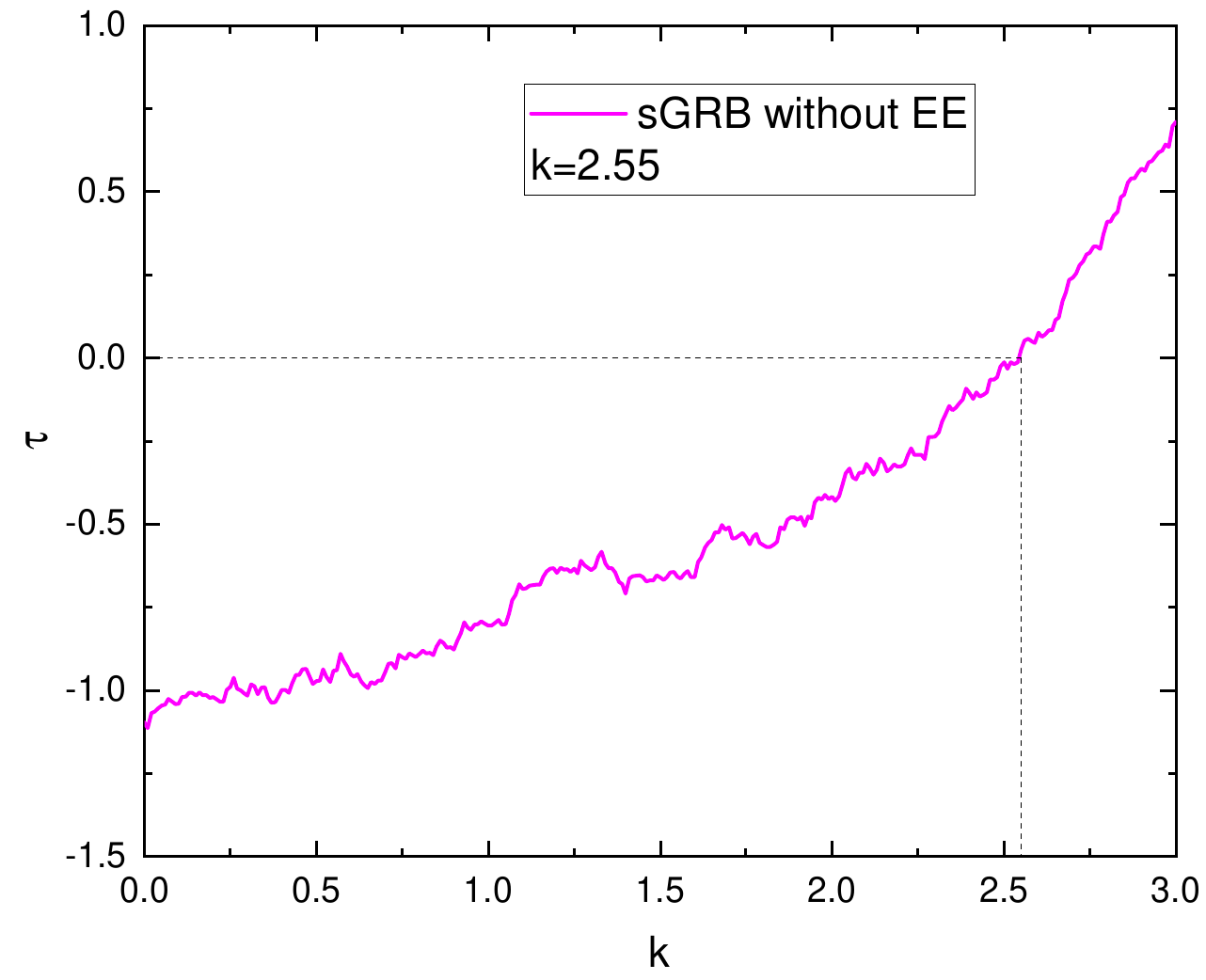}
	\includegraphics[width=\columnwidth]{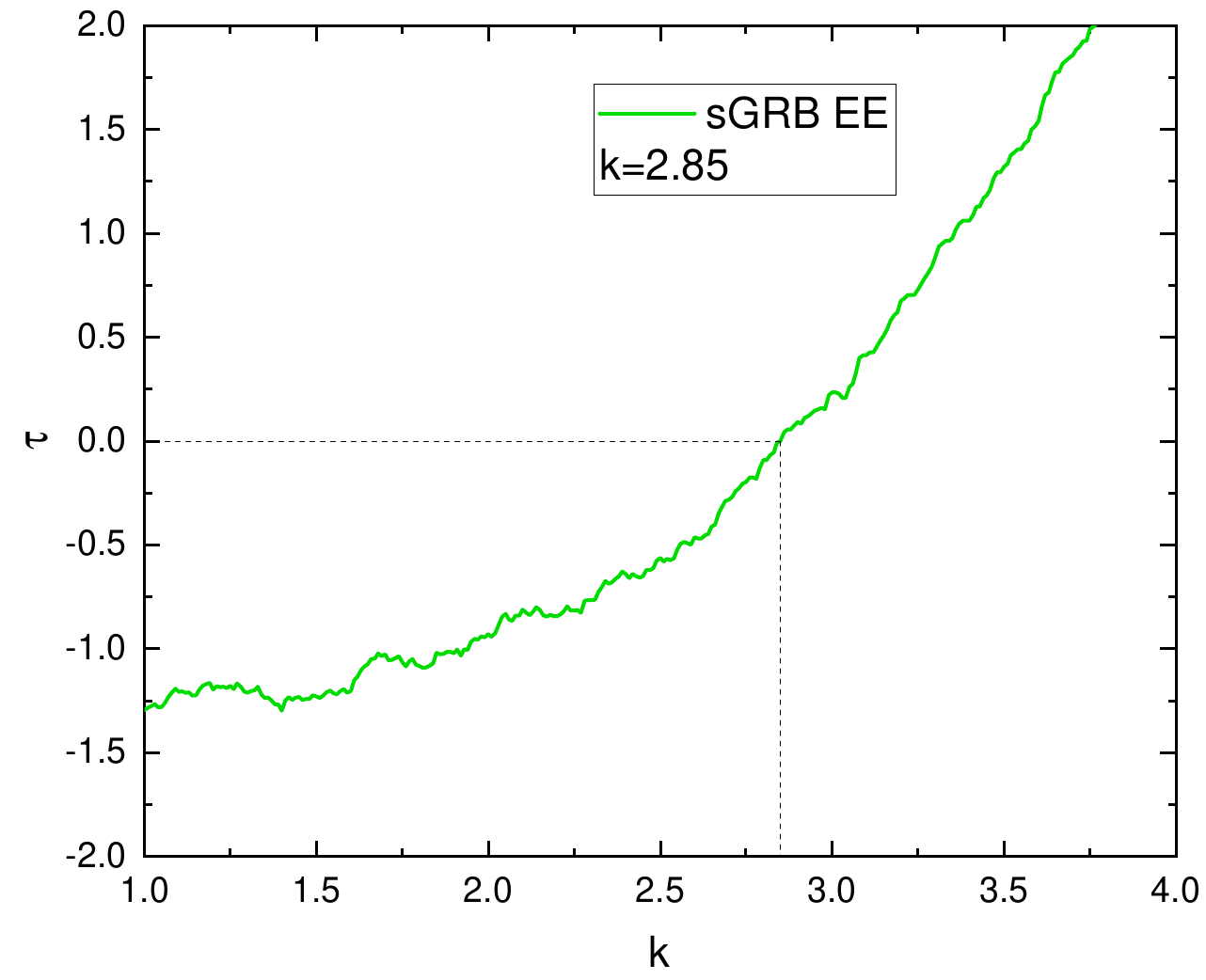}
	\caption{The black dotted line represents the best fit for $\tau=0$ at this point, we choose a form of $g(z)$ that makes test statistic $\tau=0$, then the effect of luminosity evolution can be removed by the transformation of $ L_{\rm 0} = L/g(z) $. The redshift and luminosity are independent. The value of k is 2.55 for sGRB without EE (left panel) and 2.85 for sGRB EE (right panel). }
	\label{fig:6}
\end{figure*}

\begin{figure*}
\centering
	\includegraphics[width=\columnwidth]{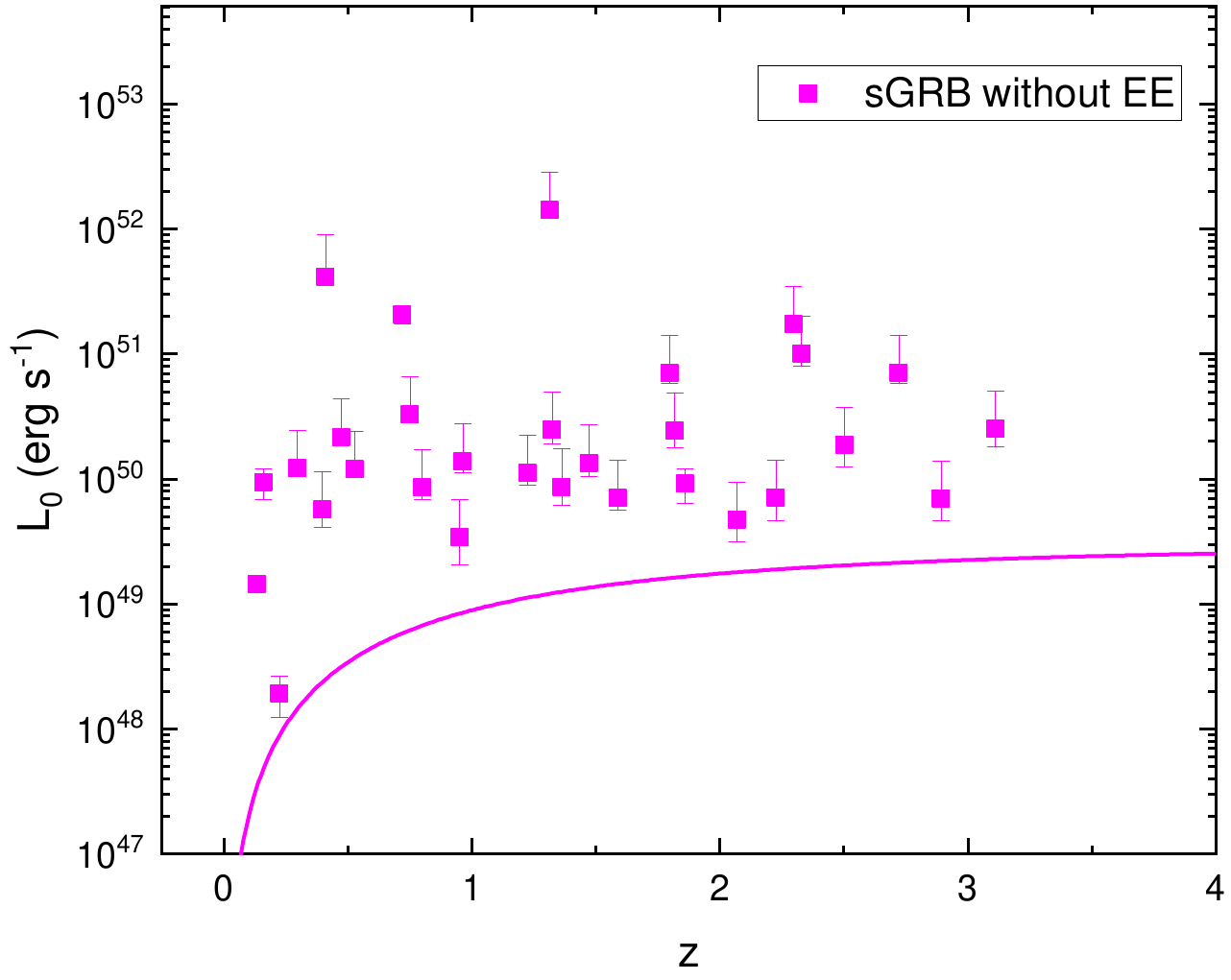}
	\includegraphics[width=\columnwidth]{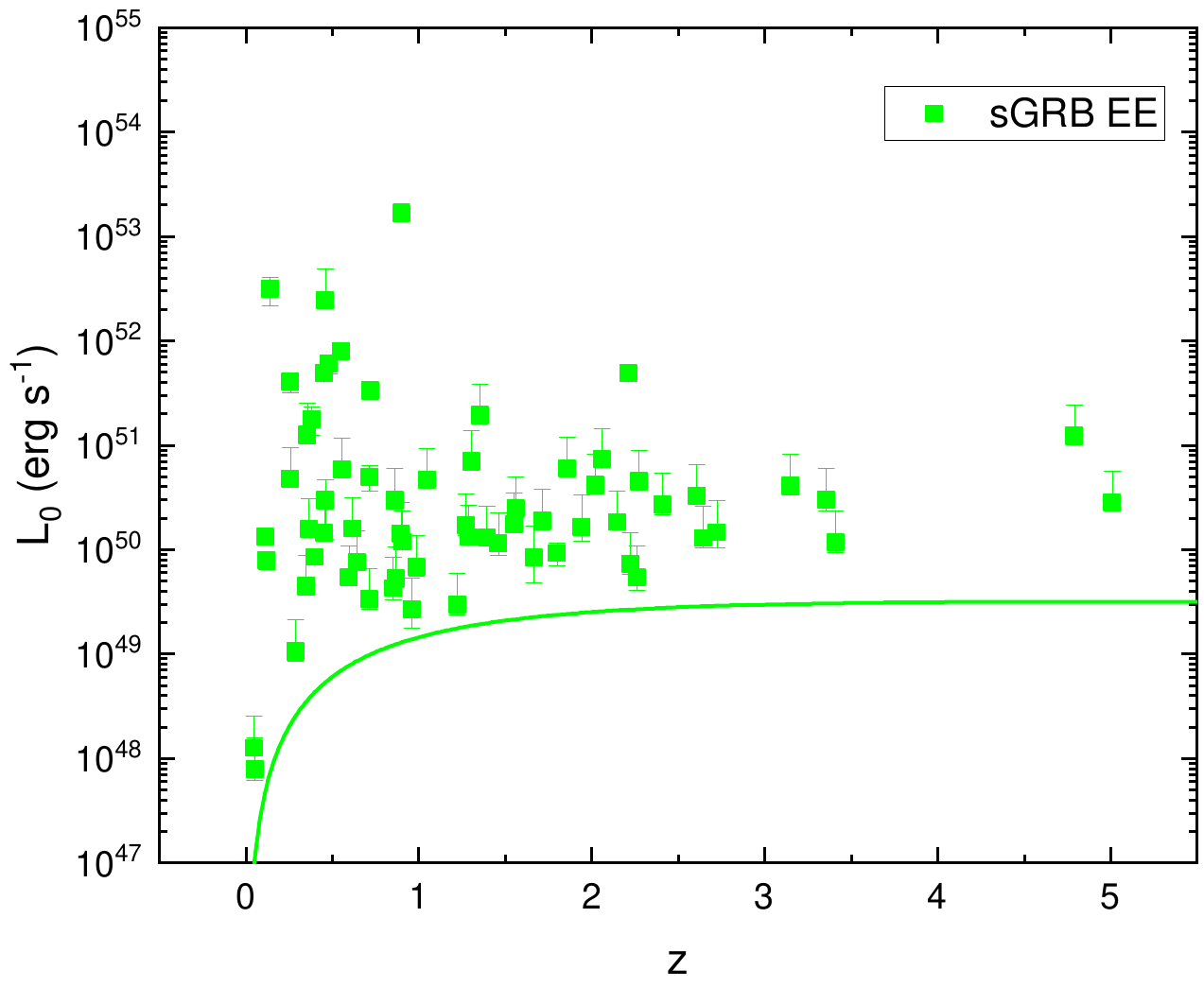}
	\caption{Redshift distribution of sGRB without EE (left panel) and sGRB EE (right panel) estimated by the $E_{p}$-$L_p$ correlation after removing co-evolution.}
	\label{fig:5}
\end{figure*}

\begin{figure*}
\centering
	
	\includegraphics[width=\columnwidth]{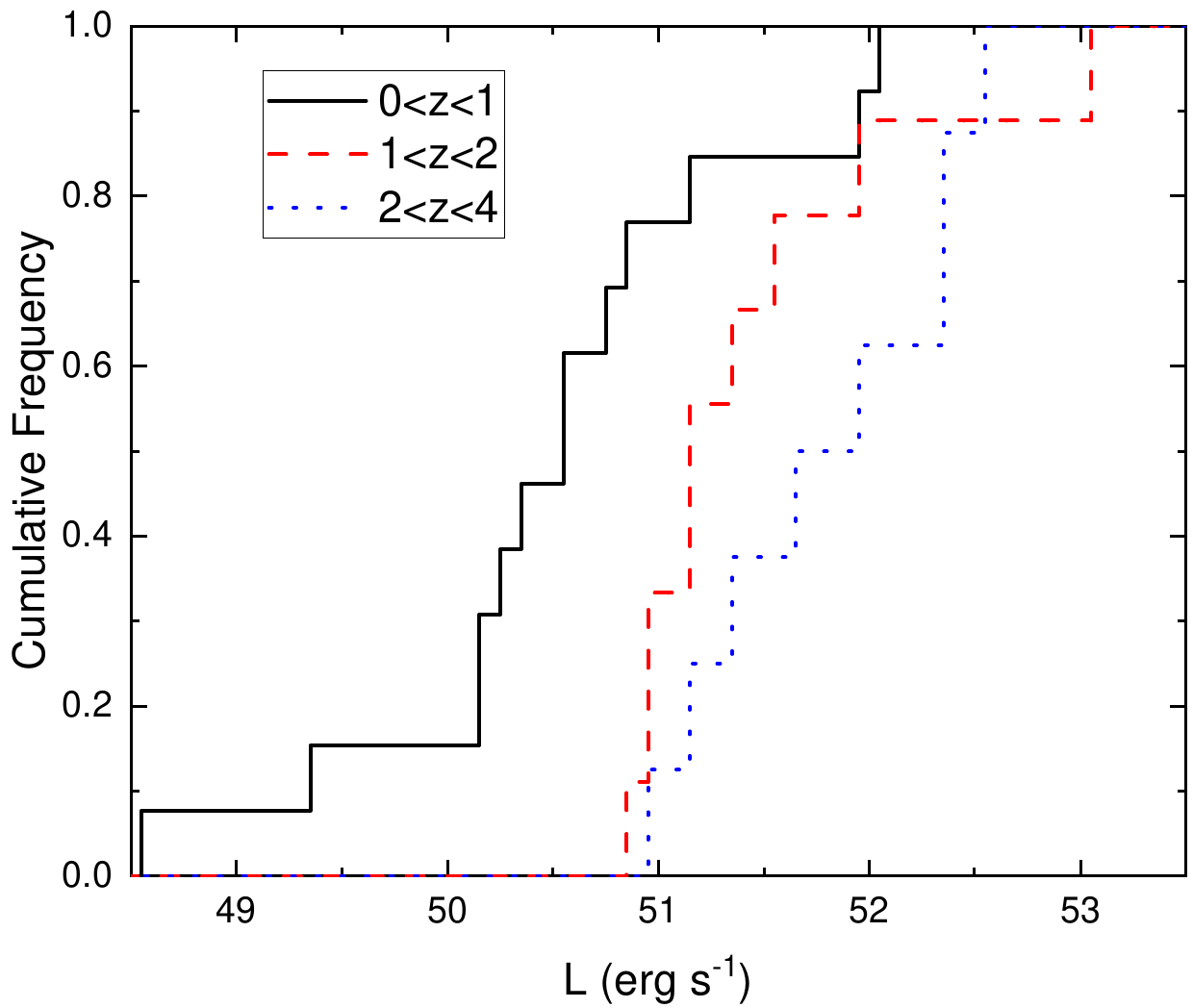}
 \includegraphics[width=\columnwidth]{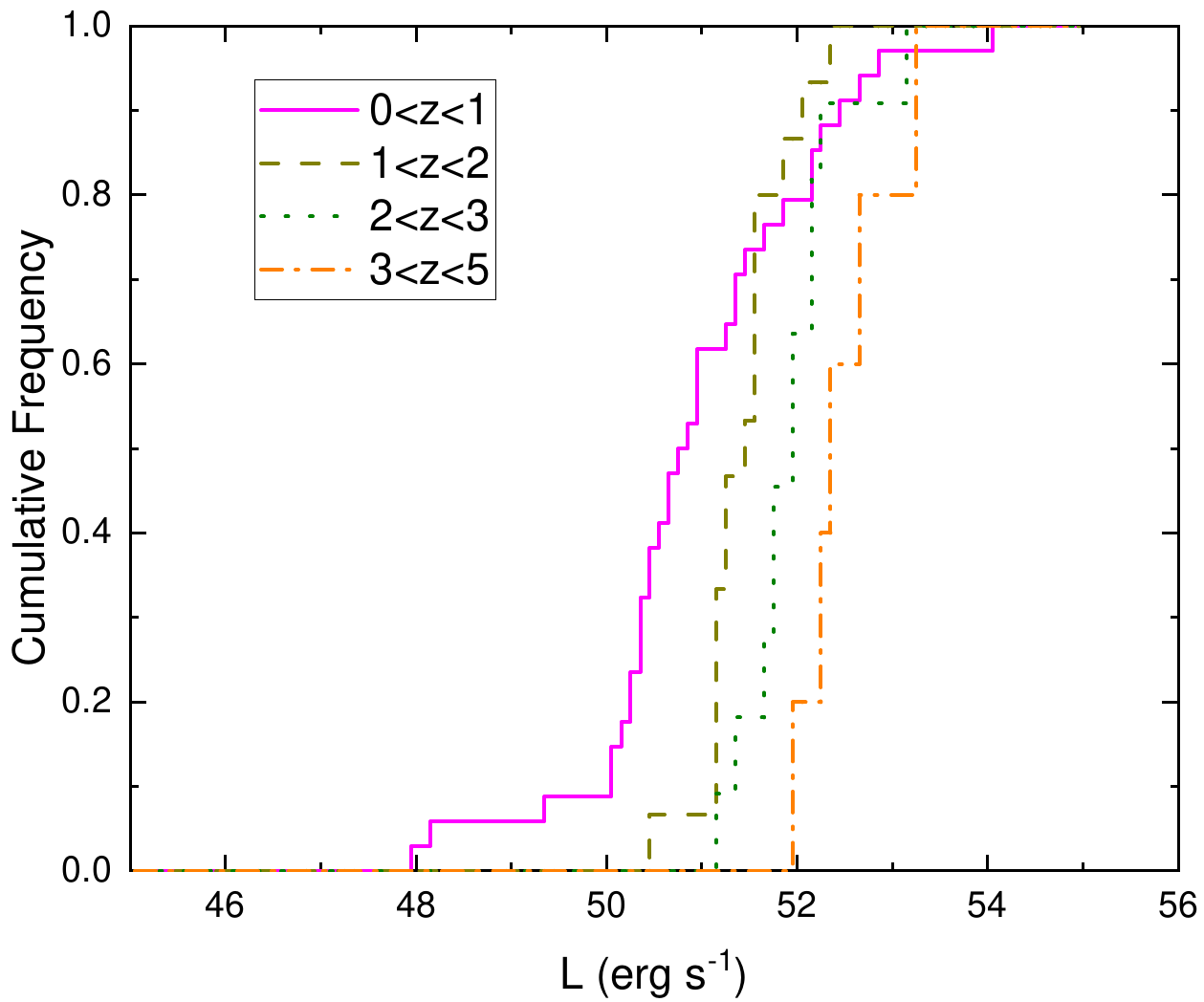}
	\caption{The cumulative luminosity distribution of sGRB without EE (left panel) and sGRB EE (right panel) in the several redshift ranges. The luminosity evolution exists because the luminosity increases toward the higher redshift.}
	\label{fig:2}
\end{figure*}

\begin{figure*}
\centering
	\includegraphics[width=\columnwidth]{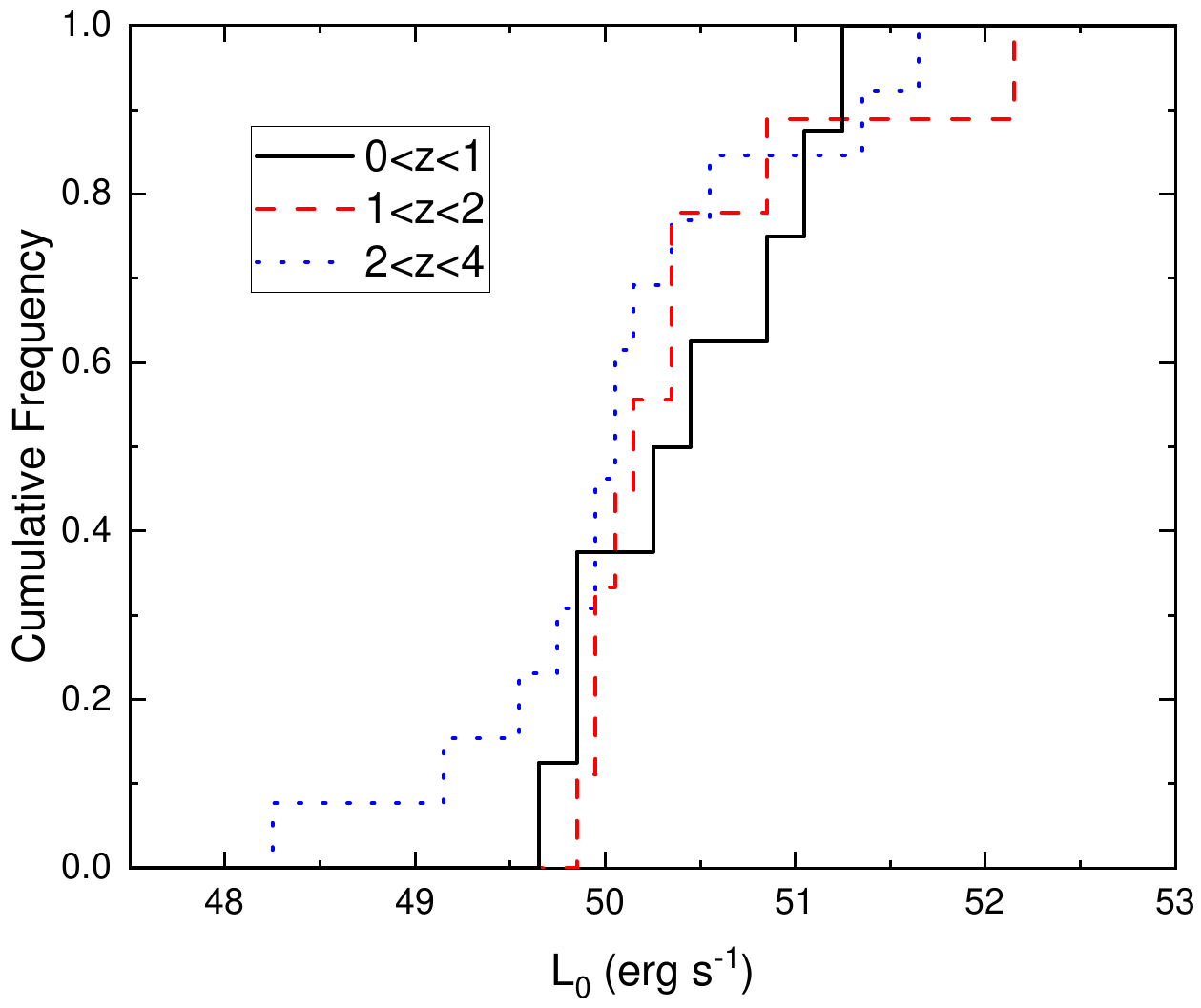}
 \includegraphics[width=\columnwidth]{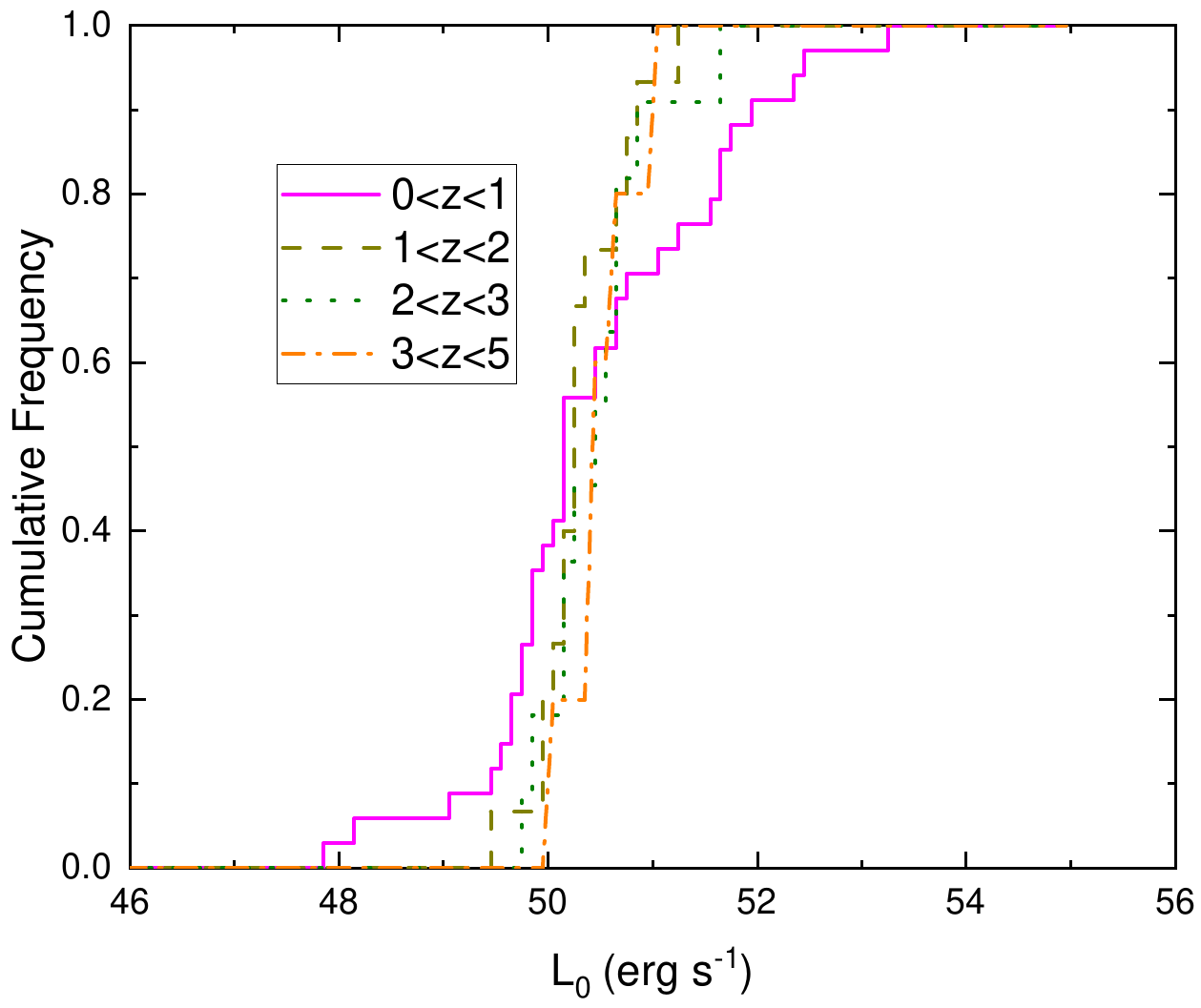}
	\caption{The cumulative luminosity distribution of sGRB without EE (left panel) and sGRB EE (right panel) in the several redshift ranges after removing co-evolution.}
	\label{fig:3}
\end{figure*}

It will be difficult to obtain the intrinsic distribution of GRBs before correcting for the selection effect (Figure \ref{fig:4}). \citet{1971MNRAS.155...95L} applied a novel approach to study the luminosity function and density evolution of a flux-limited quasar sample. The underlying assumption of Lynden-Bell's method is that the luminosity \(L\) and redshift \(z\) distributions are independent. Consequently, it is essential to eliminate this effect using the EP-L method prior to applying Lynden-Bell's approach \citep{1992ApJ...399..345E}. 
The luminosity evolves with a function ${L_{\gamma}}(z) = L_{\gamma}(1 + z)^k$, where $k$ can be determined empirically. Previous authors have used the Kendall $\tau$ statistical method to derive the value of $k$ \citep{1992ApJ...399..345E,2015ApJS..218...13Y}.

Figure \ref{fig:4} displays the distribution of luminosity and redshift. For a random point $i$ (${z_i}, L_{i}$) in this plane, ${J_{\rm{i}}}$ can be defined as
\begin{equation}
	{J_{\rm{i}}} = \{ {\rm{j}} | {L_{\rm{j}}} \ge {L_i}, {z_j} \le z_i^{\max} \},
\end{equation}
where $L_i$ is the luminosity of the $i$th sample and $z_i^{\max}$ is the maximum redshift at which the sGRB with the luminosity limit exists. This range is depicted as the black rectangle in Figure \ref{fig:4}. The number of sGRBs contained in this region is denoted as ${n_i}$. The number ${N_i} = {n_i - 1}$ represents excluding the $i$th point.

Similarly, ${J_i^{'}}$ can be defined as
\begin{equation}
	J_i^{'} = \{ {\rm{j}} | {L_{\rm{j}}} \ge L_{i}^{\lim}, {z_j} \le {z_i} \},
\end{equation}
where $L_{i}^{\lim}$ is the limiting gamma-ray luminosity at redshift ${z_i}$. In Figure \ref{fig:4}, this range is illustrated as a red rectangle. The number of samples contained within this range is denoted as ${M_i}$.

\begin{figure*}
	\centering{\includegraphics[width=\columnwidth]{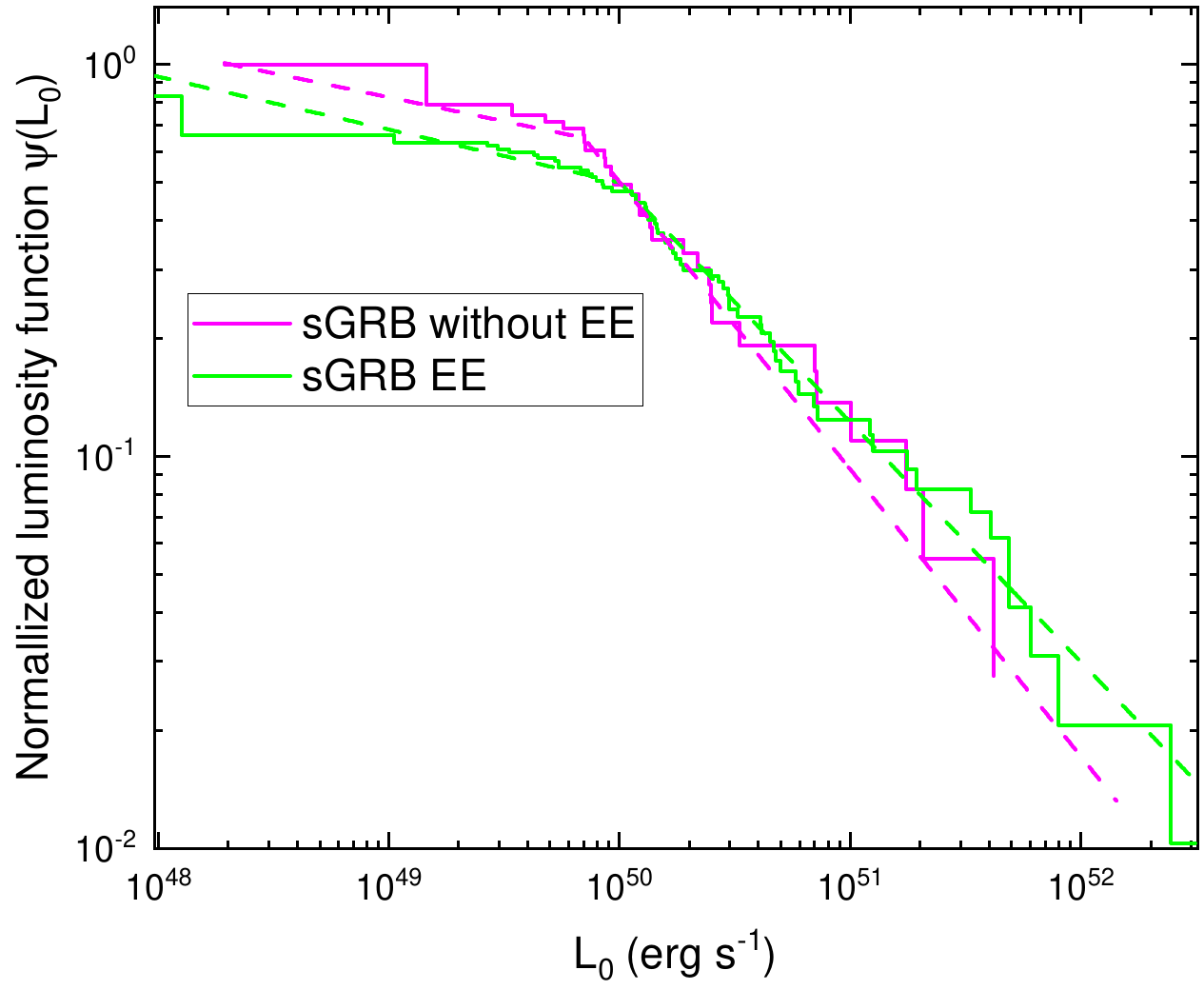}}
 \centering{\includegraphics[width=\columnwidth]{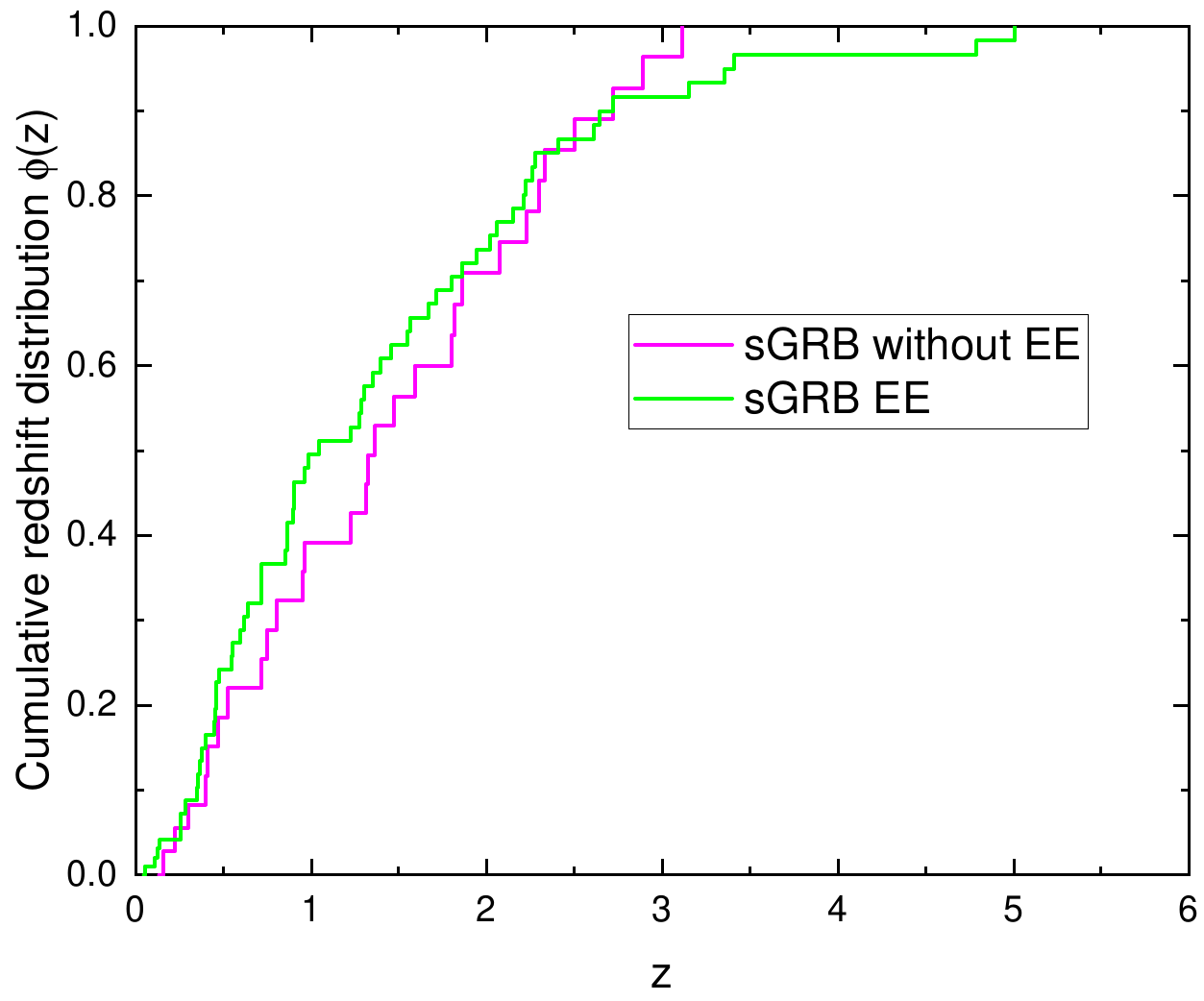}}
	\caption{Left panel: The normalized cumulative luminosity function of sGRB with (green line) and without EE (purple line). Right panel: The normalized cumulative redshift function of sGRB with  (green line) and without EE (purple line).}
	\label{fig:7_1}
\end{figure*}

Firstly, we define the number of samples that have a redshift $z$ less than or equal to ${z_i}$ in the black rectangle as ${R_i}$. The statistic $\tau$ is \citep{1992ApJ...399..345E}
\begin{equation}
	\tau = \frac{{\sum_i ({R_{\rm{i}}} - {E_i})}}{{\sqrt{\sum_i {V_{\rm{i}}}}}},
\end{equation}
where ${E_i} = \frac{{1 + n_i}}{2}$ and ${V_i} = \frac{{n_i^2 - 1}}{{12}}$ are the expected mean and variance under the hypothesis of independence, respectively. If $R_i$ is exactly uniformly distributed between 1 and $n_i$, then the samples of $R_i \geq E_i$ and
$R_i \leq E_i$ should be nearly equal, and the test statistic $\tau$ will be
nearly 0, as shown in Figrue \ref{fig:6}. The function $g(z) = (1 + z)^k$ is used to measure the correlation of the luminosity evolution function. Figure \ref{fig:5} shows the distribution of non-evolving luminosity. The values of $k$ are 2.55 and 2.85 for sGRB without EE and sGRB EE, respectively.

After correcting for the effect of luminosity evolution using \(\psi(L_0) = \psi\left(\frac{L}{1 + z}\right)^{2.55}\) for sGRB without EE and \(\psi(L_0) = \psi\left(\frac{L}{1 + z}\right)^{2.85}\) for sGRB EE, we obtain the local cumulative luminosity function \(\psi(L_0)\) using a non-parametric method based on the following equation \citep{1971MNRAS.155...95L,1992ApJ...399..345E}:
\begin{equation}
	\psi (L_{0{\rm{i}}}) = \prod_{j < i} \left( 1 + \frac{1}{{N_j}} \right),
	\label{equ:5}
\end{equation}
and the cumulative number distribution can be derived from
\begin{equation}
	\phi (z) = \prod_{j < i} \left( 1 + \frac{1}{{M_j}} \right).
	\label{equ:6}
\end{equation}

In Figure \ref{fig:2} and Figure \ref{fig:3}, we present the cumulative luminosity distribution of sGRBs with and without EE in several redshift ranges before and after removing the evolutionary effect (see Table \ref{tab:luminosity}). Luminosity evolution is evident as luminosity increases with higher redshifts. After removing the co-evolution effect, the median values of sGRB with and without EE is roughly same in different redshift range.

The differential form of $\phi (z)$ represents the formation rate $\rho $, which can be expressed as
\begin{equation}
	\rho (z) = \frac{{d\phi (z)}}{{dz}}(1 + z){\left( {\frac{{dV(z)}}{{dz}}} \right)^{ - 1}}          
	\label{equ:7}
\end{equation}
where ${\frac{{dV(z)}}{{dz}}}$ represents the differential comoving volume, which can be written as
\begin{eqnarray}
	\begin{split}
		\frac{{dV(z)}}{{dz}} =4\pi \left( {\left. {\frac{c}{{{H_0}}}} \right)} \right.\frac{{D_L^2}}{{{{(1 + z)}^2}}} \times \frac{1}{{\sqrt {1 - {\Omega _m} + {\Omega _m}{{(1 + z)}^3}} }}
		\label{equ:8}
	\end{split}
\end{eqnarray}

\section{luminosity function and formation rate of samples} \label{sec:luminosityfunction}
In this section, we respectively present results for the luminosity and formation rate of the sample. As discussed above, we obtain the results by using the non-parametric method of Kendall $\tau$ statistic test.

\subsection{luminosity function} \label{sec:luminosity}
We use broken power law to fit the cumulative luminosity function in Figure \ref{fig:7_1}. The forms of the luminosity function for sGRB without EE $\psi ({L_0})$ can be expressed as
\begin{eqnarray}
	\psi ({L_0}) \propto \left\{ {\begin{array}{*{20}{c}}
			{L_0^{ -0.12\pm0.01},{L_0} < L_0^b}\\
			{L_0^{ -0.73\pm0.02},{L_0} > L_0^b}                  
	\end{array}} \right.
	\label{equ:9}
\end{eqnarray}\\
and for sGRB EE as
\begin{eqnarray}
	\psi ({L_0}) \propto \left\{ {\begin{array}{*{20}{c}}
			{L_0^{ -0.13\pm0.003},{L_0} < L_0^b}\\
			{L_0^{ -0.61\pm0.01},{L_0} > L_0^b}                  
	\end{array}} \right.
	\label{equ:9}
\end{eqnarray}\\
where $L_0^b $ is the break point, the values of sGRB without EE and sGRB EE are ${\rm{7}} \times {\rm{1}}{{\rm{0}}^{{\rm{49}}}}{\rm{erg/s}}$ and ${\rm{1}} \times {\rm{1}}{{\rm{0}}^{{\rm{50}}}}{\rm{erg/s}}$. It is worth pointing out that the $\psi (L/g(z))$ is local luminosity at $z=0$ for the luminosity evolution is removed, and the luminosity function can be rewritten as $\psi ({L_0}) = \psi {(L/(1 + z)^{k})}$. Therefore, the break luminosity at z can be deduced $L_z^b = L_0^b{(1 + z)^{k}}$. The luminosity function of sGRB without EE is marginally consistent with sGRB EE. Our result is roughly consistent with \citet{2014ApJ...789...65Y}, They used \textit{Swift} sGRB and obtain the power law index is $ -0.84^{+0.07}_{-0.09} $. Our result is different with \citet{2018ApJ...852....1Z,2021RAA....21..254L} for bright 
\textit{fermi} sGRBs. They have more steep decline trend $\psi ({L_0}) \propto L_0^{ - 1.07}$. This may be due to the selection effect between different instruments.

\subsection{Formation rate} \label{sec:formation rate}
\begin{figure}
	\centering{\includegraphics[width=\columnwidth]{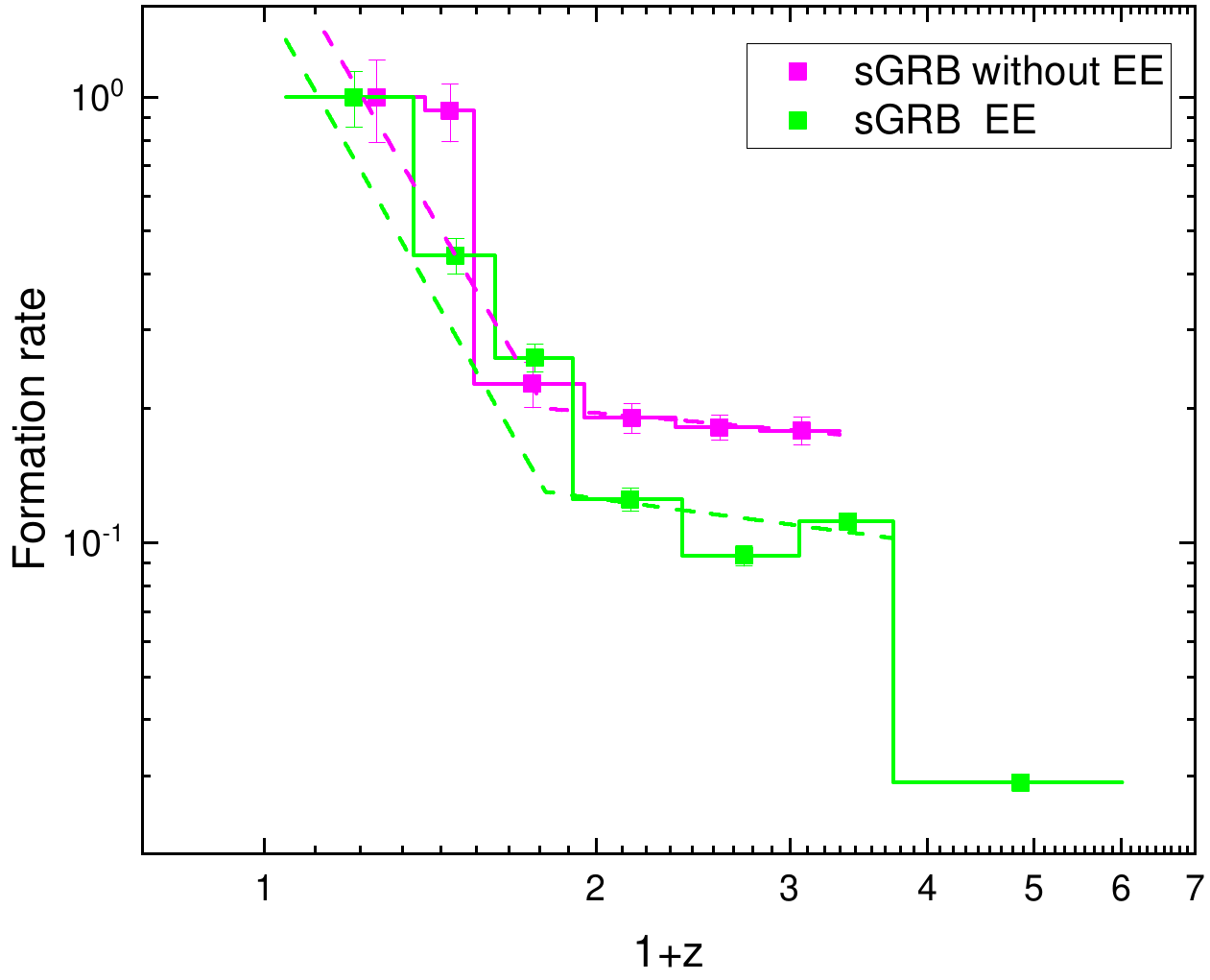}}
	\caption{The formaton rate of sGRB with (green data) and without EE (purple data).}
	\label{fig:8}
\end{figure}

Figure \ref{fig:7_1} gives the distribution of cumulative redshifts $\phi (z)$. Our main interest is the formation rate of two types sGRBs. We can calculate the formation rate $\rho (z)$ according to Equation (\ref{equ:7}). As shown in Figure \ref{fig:8}, the formation rate decreases as the redshift $z<z_0$. The best fitting broken power-law for various segments for sGRB without EE as
\begin{eqnarray}
	\rho (z) = \left\{ {\begin{array}{*{20}{c}}
			{{{(1 + z)}^{{-3.04\pm0.10}}},z < {z_0}}\\
			{{{(1 + z)}^{{-0.29\pm0.38}}},z > {z_0}}
	\end{array}} \right.
	\label{equ:10}                        
\end{eqnarray}\\
and for sGRB EE as
\begin{eqnarray}
	\rho (z) = \left\{ {\begin{array}{*{20}{c}}
			{{{(1 + z)}^{{-3.85\pm0.15}}},z < {z_0}}\\
			{{{(1 + z)}^{{-0.40\pm1.11}}},z > {z_0}}
	\end{array}} \right.
	\label{equ:10}                        
\end{eqnarray}\\
where the break point is $z_0$= 1 for sGRB without EE and sGRB EE. We found the FR of two kinds of sGRB have similar power law indexes. The formation rate $\rho (z)$ decreases at $z<1$ but keeps stable at 1< z< 3, then continues to decline. Notably, we fitted the data until $z<3$ due to this being the upper limit of sGRB. However, this operation will not affect the results, as our main comparison is the FR similarity between sGRB without EE and sGRB EE. \citet{2014ApJ...789...65Y} found that the FR can be expressed as \(\alpha = 6.0 \pm 1.7\), \(\beta = \text{constant}\), with a break redshift of \(z_0 = 1.67\), using 53 sGRBs from \textit{BATSE}, \textit{HETE-2}, and \textit{Swift/BAT}. This FR is similar to ours for \(z > 1.67\). The discrepancy for \(z < 1.67\) may be attributed to their omission of the differential comoving volume term. \cite{2018ApJ...852....1Z} and \cite{2021RAA....21..254L} reported that the sGRB FR declines rapidly towards the end. \citet{2018ApJ...852....1Z}, who selected 284 sGRBs observed by Fermi/GBM, provided the best-fitting result for the FR of sGRBs described by \(\alpha = 3.08 \pm 0.06\), \(\beta = 4.98 \pm 0.03\), with a break redshift of \(z_0 = 1.60\). Furthermore, \citet{2021RAA....21..254L} analyzed 334 sGRBs from Fermi and found the FR can be expressed as \(\alpha = -4.02 \pm 1.34\), \(\beta = 4.93 \pm 0.30\), with the break redshift \(z_m = 0.4 \pm 0.10\). This result is roughly consistent with our findings for \(z < 1\). However, for \(2 < z < 4\), our results differ from theirs, which may be due to the limited sample size within this range. The observation band of the \textit{Fermi} telescope is broader than that of \textit{Swift}, making it easier to detect high-brightness sGRBs.

\section{CONCLUSION AND DISCUSSION} \label{sec:DISCUSSION}

In this paper, we use 26 \textit{Swift} sGRBs with measured redshifts to fit the \(E_p - L_p\) correlation. Utilizing this correlation, we obtain the pseudo redshifts of 69 sGRBs observed by \textit{Swift}. Our final sample comprises 30 sGRBs without early emission (EE) and 65 sGRBs with EE, with their details listed in Tables \ref{tab:1} and \ref{tab:2}. We then apply the Lynden-Bell c$^-$ method to study the luminosity function and formation rate of sGRBs. The effect of luminosity evolution is corrected using \(L_{0} = \frac{L}{{(1 + z)^{k}}}\), where \(k\) is 2.55 for sGRB without EE and 2.85 for sGRBs with EE.

After removing the effect of luminosity evolution, we compare the luminosity functions and formation rates of sGRBs with and without EE without any assumptions. The fitting functions can be expressed as \(\psi(L_{0}) \propto L_{0}^{-0.12 \pm 0.01}\) for \(L_{0} < L_0^b\) and \(\psi(L_{0}) \propto L_{0}^{-0.73 \pm 0.02}\) for \(L_{0} > L_0^b\) for sGRB without EE. For sGRBs with EE, the fitting functions are \(\psi(L_{0}) \propto L_{0}^{-0.13 \pm 0.003}\) for \(L_{0} < L_0^b\) and \(\psi(L_{0}) \propto L_{0}^{-0.61 \pm 0.01}\) for \(L_{0} > L_0^b\). The formation rates can be described as \(\rho(z) \propto (1 + z)^{-3.04 \pm 0.10}\) for \(z < 1\) and \(\rho(z) \propto (1 + z)^{-0.29 \pm 0.38}\) for \(1 < z < 3\) for sGRB without EE, while for sGRBs with EE, the rates are \(\rho(z) \propto (1 + z)^{-3.85 \pm 0.15}\) for \(z < 1\) and \(\rho(z) \propto (1 + z)^{-0.40 \pm 1.11}\) for \(1 < z < 3\).
It is important to note that our statistical results are somewhat speculative due to the limited number of high-redshift sGRB samples. Additionally, we have analyzed the offsets of these two types of samples, and the results indicate that they follow the same distribution. There is no significant evidence suggesting that these two types of outbursts have different progenitors. Based on these findings, we propose that the presence or absence of EE may not be a reliable indicator of whether the sample originates from BNS merger or NS-BH merger.

There are other ways to distinguish whether a GRB originates from the merger of NS-BH. Detection of the electromagnetic counterparts of gravitational wave sources and GRBs is important to unveil the nature of compact binary coalescences.
Theoretically, an NS-BH merger must result in a BH central engine, whereas a BNS merger could result in an NS remnant \citep{2006PhRvD..73f4027S}. GRBs could be powered by newly born magnetars' spin-down \citep{1998A&A...333L..87D,2011MNRAS.413.2031M}.
\citet{2023ApJ...955...98L} suggested that sGRB 130310A could be a probable candidate originating from an NS-BH merger based on the existence of Quasi-Periodic Oscillations (QPO).
The merger of compact binary stars generates gravitational waves (e.g., GW 170817 has been confirmed to originate from the merger of binary neutron stars), and events like S190814bv \citep{2019GCN.25324....1L}, GW 200105, and GW 200115 were classified as NS-BH mergers \citep{2023PhRvD.108h3018Y}. Therefore, the volumetric rates of NS-BH and BNS mergers are expected to be distinguishable.
More material is ejected during NS-BH mergers, resulting in brighter kilonovae \citep{2016ApJ...825...52K,gompertz2023multimessenger}. NS-BH kilonovae may also appear redder due to their ejecta retaining a high neutron fraction and being enriched in lanthanide elements \citep{2014ApJ...780...31T}. This provides a way to identify the origin of GRBs in the future by comparing the properties of kilonovae, not limited to sGRB samples, as a portion of lGRBs has been confirmed to be associated with kilonovae \citep{2022Natur.612..232Y,2024Natur.626..742Y}.

The EE is defined as a low-intensity burst following the initial main emission \citep{2006ApJ...643..266N}. The merger of NS-BHs generates more energy and takes longer than BNS, so EE radiation may be an indicator of their origin \citep{2020ApJ...895...58G}. However, our results found that these two types of bursts have the same distribution of luminosity function and formation rate, implying no obvious difference from the perspective of formation rate. The generation of EE radiation may come from other reasons. 
Spin-down of a strongly magnetized neutron star by losing rotational energy \citep{2012MNRAS.419.1537B,2013MNRAS.430.1061R,2014MNRAS.438..240G}.
EE generated by relativistic winds, which are the result of the formation and early evolution of highly magnetized, rapidly rotating neutron stars \citep{2008MNRAS.385.1455M,2022ApJ...939..106J}. Material fallback heated by r-process \citep{2019MNRAS.485.4404D,2022ApJ...936L..10Z,2024ApJ...966L..31M}.


\section*{Acknowledgements}

	This work was supported by the basic research project of Yunnan Province (Grant No. 202301AT070352). In this paper, we use data from the \textit{Swift}.
All of the data presented in this article were obtained from the \textit{Swift}/BAT \url{https://swift.gsfc.nasa.gov/results/batgrbcat/},\url{https://swift.gsfc.nasa.gov/archive/grb_table/} and table 1 of \citet{2021RAA....21..254L}. We would like to extend our sincere gratitude to the reviewers for their thorough evaluations and constructive comments, which significantly improved the quality of our manuscript. Our heartfelt thanks also go to Brendan O'Connor from the Department of Physics at George Washington University and Fayin Wang from the School of Astronomy and Space Science at Nanjing University for their invaluable suggestions and support throughout the submission process. Additionally, we are deeply appreciative of Siyuan Zhu from the School of Physics and Astronomy at Sun Yat-sen University for his expert guidance on the Bayesian Block method code.

\bibliography{sample631}{}
\bibliographystyle{aasjournal}

\section{Appendix}\label{sec:Appendix}
The appendix includes the Tables \ref{tab:1}, \ref{tab:2} and \ref{tab:luminosity}. Tables \ref{tab:1} and \ref{tab:2} show the parameters of short gamma ray bursts without and with extended emission, respectively. Tables \ref{tab:luminosity} presents the median luminosity values for sGRBs with and without EE across different redshift range before and after removing co-evolution effect.

\renewcommand\thetable{A1}
\begin{deluxetable*}{lccccccc}
\tablenum{A1}
\tablecaption{Parameters of short gamma ray bursts without extended emission.\label{tab:1}}
\tablehead{
\colhead{Name} &	\colhead{$T_{90}$ (s)} & \colhead{z} &   \colhead{$alpha$} &\colhead{$E_p$ (keV)} & 
	\colhead{$F_p$ ($photon/c{m^2}/s$)}  & \colhead{$L_p$ ($erg/s$)}  & \colhead{offset (kpc)}}
\startdata
	050202&0.27&1.86&-0.32&84.26&0.55&1.34E+51&55.19 \\
050509B&0.07&0.23&2.19&58.85&0.28&3.26E+48&63.70 \\
050925&0.07&0.30&0.11&64.80&10.00&2.37E+50&-\\
070209&0.09&2.23&0.56&81.38&0.38&1.41E+51&-\\
071112B&0.30&1.32&0.72&124.38&1.30&2.12E+51&-\\
080426&1.70&0.47&-1.75&69.80&4.80&5.86E+50&-\\
080702A&0.50&3.11&-0.89&157.49&0.70&9.28E+51&-\\
081226A&0.40&1.80&-1.21&210.06&2.40&9.69E+51&$<4.06$\\
090305A&0.40&2.33&-0.55&255.79&1.90&2.17E+52&-\\
090417A&0.07&0.40&0.39&42.78&3.60&1.34E+50&-\\
090621B&0.14&0.75&-0.29&122.48&3.90&1.37E+51&-\\
090815C&0.60&2.50&0.10&131.18&0.60&4.61E+51&-\\
100206A$^*$&0.12&0.41&-0.40&638.98&1.40&1.00E+52&25.28 \\
100628A&0.04&1.59&2.66&74.33&0.50&8.00E+50&-\\
101224A$^*$&0.20&0.72&0.93&566.94&0.70&8.24E+51&12.75 \\
110112A&0.50&0.95&-0.89&33.69&0.50&1.88E+50&-\\
110420B&0.08&1.36&1.64&78.30&0.70&7.78E+50&-\\
130626A&0.16&1.47&-0.17&95.80&0.90&1.36E+51&-\\
150101A&0.06&2.89&-1.00&80.00&0.30&2.23E+51&-\\
150101B$^*$&0.02&0.13&-1.36&125.11&0.37$^1$&2.00E+49&7.36 \\
160411A&0.36&0.80&1.40&66.59&1.50&3.87E+50&1.40 \\
160714A&0.35&1.82&-0.98&129.09&1.10&3.42E+51&-\\
160726A&0.70&2.30&-0.96&333.08&3.20&3.66E+52&-\\
160821B$^*$&0.48&0.16&-1.40&97.44&1.70&1.38E+50&15.75 \\
170112A&0.06&2.07&3.48&65.23&0.30&8.26E+50&-\\
180727A&1.10&0.53&-0.59&67.63&3.60&3.55E+50&2.56 \\
190326A&0.08&0.96&-1.00&80.00&1.55&7.74E+50&-\\
190427A&0.30&2.72&-1.05&240.15&1.60&2.03E+52&-\\
210119A&0.06&1.22&0.06&83.49&1.00&8.62E+50&-\\
230217A&1.30&1.31&-0.87&657.79&20.20&1.21E+53&-\\
\enddata
\begin{threeparttable} 
	\begin{tablenotes} 
		\item [*] The redshift of these samples are acquired from Swift/BAT network and \citet{2021RAA....21..254L}.\\ other sample's redshift is estimated by $E_{p,i}-L_p$ relation.
  \item [1] The unit of $F_p$ is $erg/c{m^2}/s$
	\end{tablenotes} 
\end{threeparttable}		
\end{deluxetable*}

\renewcommand\thetable{A2}
\setlength{\tabcolsep}{1pt}
\begin{longtable*}{lccccccc}
	\caption{Parameters of short gamma ray bursts with extended emission.}\label{tab:2}\\
	\toprule
	Name&$T_{90} (s)$&z&$\alpha$&$E_p$ (keV) &$F_p$ ($photon/c{m^2}/s$) &$L_p$ ($erg/s$)&offset (kpc)\\
	\midrule
	\endfirsthead
	\multicolumn{8}{c}
	{{\bfseries{Table \thetable\ }continued from previous page}} \\
	\toprule
Name&$T_{90} (s)$&z&$\alpha$&$E_p$ (keV) &$F_p$ ($photon/c{m^2}/s$) &$L_p$ ($erg/s$)&offset (kpc)\\

	\midrule
	\endhead
	\midrule
	\multicolumn{8}{c}
	{{ }} \\
	\endfoot
	\endlastfoot
050813$^*$&0.45 &1.80 &1.42 &64.62 &0.94 &1.74E+51&43.57\\
051210&1.30 &5.01 &-0.75 &243.93 &0.75 &4.65E+52&29.08\\
051221A$^*$&1.40 &0.55 &-1.20 &621.69 &12.00 &2.77E+52&2.08 \\
060502B$^*$&0.13 &0.29 &-0.11 &117.95 &0.62 &2.17E+49&70.00 \\
060801&0.49 &2.02 &0.28 &190.29 &1.27 &9.63E+51&10.25\\
061201$^*$&0.76 &0.11 &-0.72 &872.00 &3.86 &1.00E+49&32.47\\
070429B$^*$&0.47 &0.90 &-1.10 &72.85 &1.76 &7.40E+50&6.00\\
070714A&2.00 &0.05 &-1.97 &1.81 &1.80 &1.43E+48&-\\
070724A$^*$&0.40 &0.46 &-0.68 &56.49 &1.00 &5.00E+50&5.52\\
070729&0.90 &2.41 &-0.32 &174.63 &1.00 &8.84E+51&19.72\\
070809&1.30 &1.55 &-1.43 &111.13 &1.20 &2.52E+51&34.11\\
071227$^*$&1.80 &0.38 &-0.23 &1000.00 &1.60 &1.39E+51&15.50\\
080905A$^*$&1.00 &0.12 &-0.63 &349.71 &1.30 &2.00E+49&14.74\\
081101&0.20 &0.62 &0.25 &91.24 &3.60 &6.24E+50&-\\
090426$^*$&1.20 &2.61 &-1.11 &55.09 &2.40 &1.26E+52&0.49\\
090510$^*$&0.30 &0.90 &-0.86&8679.58 &9.70 &2.40E+53&10.51\\
090515&0.04 &0.45 &-0.15 &79.76 &5.70 &4.20E+50&79.16\\
091109B&0.30 &1.35 &-0.40 &310.47 &5.40 &2.22E+52&4.22\\
100117A$^*$&0.30 &0.26 &-0.10&325.43 &2.90 &1.74E+51&1.35\\
100625A$^*$&0.33 &0.45 &-0.59&739.36 &2.60 &6.00E+50&2.63\\
100702A&0.16 &0.90 &-1.06 &87.50 &2.00 &8.81E+50&-\\
100724A$^*$&1.40 &1.29 &-0.51 &42.50 &1.90 &1.40E+51&-\\
101219A$^*$&0.60 &0.72 &-0.62 &841.82 &4.10 &2.30E+51&5.48\\
111117A$^*$&0.47 &2.21 &-0.50&543.61 &1.35 &1.91E+52&10.52\\
120229A&0.22 &2.22 &-0.47 &94.04 &0.50 &2.04E+51&-\\
120305A&0.10 &1.86 &-0.68 &222.89 &2.20 &1.19E+52&18.09\\
120403A&1.25 &0.99 &-0.42 &64.90 &1.10 &4.81E+50&-\\
120521A&0.45 &2.15 &-0.39 &142.87 &0.90 &4.82E+51&-\\
120804A&0.81 &0.55 &-0.97 &156.11 &10.80 &2.04E+51&2.30\\
130515A&0.29 &2.27 &-0.33 &209.68 &1.40 &1.32E+52&61.22\\
130603B$^*$&0.18 &0.36 &-0.75 &997.50 &6.40 &3.00E+51&5.40\\
131004A$^*$&1.54 &0.72 &-1.23 &117.91 &3.40 &2.33E+51&0.80 \\
140129B&1.36 &0.05 &-1.99 &1.50 &0.90 &9.17E+47&1.76\\
140516A&0.19 &1.22 &0.49 &44.57 &0.50 &2.88E+50&-\\
140606A&0.34 &2.64 &1.07 &134.25 &0.50 &5.14E+51&-\\
140622A$^*$&0.13 &0.96 &1.23 &44.24 &0.60 &1.82E+50&32.95\\
140903A$^*$&0.30 &0.35 &-1.36 &44.17 &2.50 &1.05E+50&0.90\\
141212A$^*$&0.30 &0.60 &-1.15 &94.87 &1.20 &2.05E+50&18.75\\
150120A$^*$&1.20 &0.46 &-1.81 &0.46 &1.80 &7.21E+52&4.77 \\
150423A$^*$&0.22 &1.39 &0.44 &120.50 &0.90 &1.56E+51&18.75\\
151127A&0.19 &2.26 &0.18 &84.41 &0.40 &1.57E+51&-\\
151205B&1.40 &2.72 &-1.39 &132.03 &0.70 &6.19E+51&-\\
151228A&0.27 &1.46 &0.58 &101.97 &0.90 &1.47E+51&-\\
151229A&1.78 &0.37 &-1.44 &65.88 &7.20 &3.76E+50&8.16\\
160408A&0.32 &2.06 &-0.49 &246.35 &2.10 &1.75E+52&13.13\\
160525B&0.29 &0.85 &-0.60 &45.00 &0.90 &2.47E+50&0.40\\
160601A&0.12 &3.35 &-0.57 &209.37 &0.90 &1.98E+52&1.38\\
160624A$^*$&0.20 &0.48 &-0.62&1247.50 &0.50 &3.80E+51&9.63\\
170127B&0.51 &3.15 &-0.52 &230.20 &1.10 &2.35E+52&10.37\\
170428A$^*$&0.20 &0.14 &-0.47&982.00 &2.80 &1.39E+52&7.72\\
170524A&0.10 &3.41 &-0.46 &142.22 &0.50 &8.07E+51&-\\
170728A&1.25 &0.71 &1.00 &38.79 &1.10 &1.54E+50&32.25\\
180204A&1.16 &1.30 &-0.76 &212.27 &3.60 &7.49E+51&-\\
180805A&1.68 &0.87 &-0.17 &54.08 &1.10 &3.13E+50&-\\
190627A$^*$&1.60 &1.94 &-1.48 &16.47 &1.10 &3.60E+51&-\\
200411A&0.22 &1.72 &-0.23 &133.39 &1.10 &3.26E+51&41.98\\
200522A$^*$&0.62 &0.40 &-0.54 &78.00 &1.50 &2.97E+49&0.93\\
201006A&0.49 &0.64 &0.40 &62.48 &2.20 &3.12E+50&-\\
201221D$^*$&0.16 &1.05 &-0.97 &88.31 &5.60 &3.57E+51&29.35\\
210323A&1.12 &4.79 &-1.40 &511.58 &2.90 &1.82E+53&5.89\\
210726A&0.39 &1.67 &-0.91 &85.79 &0.70 &1.38E+51&12.26\\
210919A&0.16 &1.27 &-1.15 &107.60 &1.50 &1.77E+51&51.05\\
211023B$^*$&1.30 &0.86 &-1.98 &64.17 &2.20 &1.75E+51&3.84\\
220730A&0.19 &1.56 &-1.01 &142.56 &1.60 &3.64E+51&-\\
231117A$^*$&0.67 &0.26 &-1.32 &165.66 &28.90 &9.14E+50&-\\

	\bottomrule
\end{longtable*}

\newpage
\setlength{\tabcolsep}{30pt} 
\begin{deluxetable*}{ccccc}
	\tablenum{A3}
 
	\tablecaption{Median luminosity values for sGRBs with and without extended emission in different redshift ranges.\label{tab:luminosity}}
	\tablehead{
		\colhead{Types} &\colhead{$z$} & \colhead{$L(erg/s) ^{\rm a} $} & \colhead{$L_0(erg/s)^{\rm b}$}}
	\startdata
sGRB without EE&$0<z<1$&3.55$\times 10^{50}$&1.21$\times 10^{50}$\\
&$1<z<2$&1.36$\times 10^{51}$&1.35$\times 10^{50}$\\
&$2<z<4$&6.54$\times 10^{51}$&2.18$\times 10^{50}$\\
\hline
sGRB EE&$0<z<1$&6.79$\times 10^{50}$&1.50$\times 10^{50}$\\
&$1<z<2$&2.52$\times 10^{51}$&1.71$\times 10^{50}$\\
&$2<z<3$&8.84$\times 10^{51}$&2.68$\times 10^{50}$\\
&$3<z<5$&2.35$\times 10^{52}$&3.00$\times 10^{50}$\\
	\enddata
	\begin{threeparttable} 
		\begin{tablenotes} 
  		\item [a] Median luminosity before removing the co-evolution effect.
		\item [b] Median luminosity after removing the co-evolution effect.

		\end{tablenotes} 
	\end{threeparttable}		
\end{deluxetable*}

\end{document}